\documentclass{JHEP3}
\usepackage{amsmath}
\input epsf
\usepackage{epsfig}
\usepackage{amssymb}
\usepackage{graphics}
\usepackage[active]{srcltx}

\newcommand{\bea}{\begin{eqnarray}}
\newcommand{\eea}{\end{eqnarray}}
\newcommand{\bean}{\begin{eqnarray*}}
\newcommand{\eean}{\end{eqnarray*}}
\newcommand{\nn}{\nonumber \\}

\def\W #1{\widetilde{#1}}

\def\a{{\alpha}}

\def\b{{\beta}}

\def\eps{\epsilon}

\def\Label#1{\label{#1}%
  \smash{\hbox to0pt{\raise1ex\hbox{\tiny[#1]}\hss}}}

\title{Relations of loop partial amplitudes  in gauge theory by Unitarity cut method}
\author{ Bo Feng$^{abc}$, Yin Jia$^a$, Rijun Huang$^a$\footnote{The
unusual ordering of authors is just to satisfy outdated requirement
for Ph. Degree.}
 \bigskip\\
{$^a$\small Zhejiang Institute of Modern Physics, Zhejiang
University, Hangzhou, 310027, P. R. China\\$^b$\small Center of
Mathematical Science, Zhejiang University, Hangzhou, China \\
$^c$\small Kavli Institute for Theoretical Physics China, CAS,
Beijing 100190, China }}
\date{\today}
\abstract{ It is well known that under the color-decomposition,
one-loop amplitude of gluons contains partial amplitudes of single
and double trace structures, and particularly all partial amplitudes
of double trace structure can
 be expressed as a linear combination of partial amplitudes of single
 trace structure. Using unitarity cut method, we prove that
this result is the
 natural consequence of tree-level Kleiss-Kuijf relation.
 Generalizing the unitarity cut method to two-loop (triple cut in this case), we show that, unlike
the one-loop case, { partial amplitudes of double and triple trace
structures can not
 be expressed as a linear combination of partial amplitudes of leading-color single
 trace structure}. For partial amplitudes of  subleading-color single trace structure,
we have shown a very nontrivial Kleiss-Kuijf relation for six and
seven-point amplitudes, which is one new result of our paper and can
not be obtained by $U(1)$-decoupling method. Mysteriously, when we
consider the case of eight points, Kleiss-Kuijf relation must be
modified for subleading-color single trace partial amplitudes.

}

\begin{document}

%%%%%%%%%%%%%%%%%%%%%%
\section{Introduction}
%%%%%%%%%%%%%%%%%%%%%%

Great effects have been made to explore the hidden simplicity of
amplitudes in recent years. Different from traditional lagrangian
description,  new approaches, such as the unitarity cut method
\cite{Landau:1959fi,Bern:1994cg,Bern:1994zx,Britto:2004nc,Britto:2005ha,Anastasiou:2006jv},
the Cachazo-Svrcek-Witten (CSW) rule \cite{Cachazo:2004kj} based on
the twistor string theory proposed by Witten \cite{Witten:2003nn}
and the Britto-Cachazo-Feng-Witten (BCFW) on-shell recursion
relation \cite{Britto:2004ap, Britto:2005fq}%\footnote{The
%generalization of BCFW recursion relation to supersymmetric theory
%has also been discussed in
%\cite{Bianchi:2008pu,Brandhuber:2008pf,ArkaniHamed:2008gz,Elvang:2008na},
%which helps to construct all the tree amplitudes of $\mathcal{N}=4$
%super-Yang-Mills theory \cite{Drummond:2008cr}.}
, have shown their
advantages in the calculation of scattering amplitudes. These new
methods are not only useful for calculation, but also for
understanding many properties of amplitudes.

For example,
 the well known
color-reflection, $U(1)$-decoupling
\cite{Mangano:1987xk,Bern:1990ux,Mangano:1990by} and Kleiss-Kuijf
(KK) relation \cite{Kleiss:1988ne}\footnote{The KK relation has
been proved by field theory method in \cite{DelDuca:1999rs} and by
string theory method in \cite{BjerrumBohr:2009rd,Stieberger:2009hq}.
}, together with the newly discovered Bern-Carrasco-Johansson (BCJ)
relation \cite{Bern:2008qj} for color-ordered tree-level partial
amplitudes of gauge theory, have been re-understood from the point
of view of pure field theory in \cite{Feng:2010my, Chen:2011jx} (See
further generalization and discussion
\cite{Jia:2010nz,Huang:2010fc,Zhang:2010ve}). The tree-level partial
amplitudes are defined by the color-decomposition of full tree-level
amplitude  based on its color trace structure $T^a$
\cite{Mangano:1987xk,Berends:1987cv,Mangano:1988kk,Dixon:1996wi} (or
structure constant $f^{abc}$ \cite{DelDuca:1999rs}) as follows
\bea {\cal A}^{full}_n(\{k_i,\lambda_i,a_i\})=g^{n-2}\sum_{\sigma\in
S_n/Z_n}\mbox{Tr}(T^{a_{\sigma_1}}\cdots T^{a_{\sigma_n}})
A_n(k_{\sigma_1}^{\lambda_{\sigma_1}},\ldots,k_{\sigma_n}^{\lambda_{\sigma_n}})~,
~~~\label{Tree-dec}\eea
where $k_i,\lambda_i,a_i$ are respectively momentum, helicity and
color index of $i$-th external gluon, and $S_n/Z_n$ represents the
permutations  on $n$-particles up to  cyclic ordering. The
decomposition (\ref{Tree-dec}) has separated the dynamical
information (given by the partial amplitudes) from the group
information (given by trace structure).

For tree-level partial amplitudes $A_{n}$ defined in
(\ref{Tree-dec}), because  the cyclic invariant of trace, there are
$(n-1)!$ partial amplitudes. However not all of these partial
amplitudes are algebraic independent and  they are  related to each
other by cross symmetry and other considerations.  First there is
nontrivial KK relation \cite{Kleiss:1988ne} given by
\bea A_n(1,\alpha,n,\beta)=(-)^{n_\beta}\sum_{\sigma\in
OP\{\alpha\}\cup\{\beta^T\}}A_n(1,\sigma,n)~,~~~\label{KK}\eea
where $n_\b$ is the number of $\b$-set and the Order-Preserved(OP) sum
is over all permutations of the set $\a \bigcup \b^T$ where the
relative ordering in each set  $\a$  and $\b^T$ (which is the
reversed ordering of set $\b$) is preserved\footnote{One non-trivial
example of KK relation with six gluons is given as follows
\bea A(1,2,3,6,4,5) & = & A(1,2,3,5,4,6)+ A(1,2,5,3,4,6)+
A(1,2,5,4,3,6)\nn & & +
A(1,5,4,2,3,6)+A(1,5,2,4,3,6)+A(1,5,2,3,4,6)~.~~~ \eea
}. The KK relation will reduce the number of independent partial
amplitudes to $(n-2)!$. Beyond the KK relation, there is also BCJ
relation \cite{Bern:2008qj}, which further reduces the number of
independent partial amplitudes to $(n-3)!$, where kinematic factors
$s_{ij}=(k_i+k_j)^2$ show up in the coefficients of the relation.  The
BCJ relation was originally observed from non-trivial Jacobi
relations between $s,t,u$-channels of Feynman diagrams, and then
proved  as the imaginary part of monodromy relations in string
theory \cite{BjerrumBohr:2009rd,Stieberger:2009hq}. This relation
has been proved by BCFW recursion relation in pure field theory
\cite{Feng:2010my,Chen:2011jx}\footnote{ An extension of BCJ
relation to matter fields can be found in \cite{Sondergaard:2009za}.
Other related works see
\cite{Tye:2010dd,BjerrumBohr:2010zs,Mafra:2009bz,Mafra:2010jq,BjerrumBohr:2011xe}.
}.

Beyond tree-level amplitude, a similar color decomposition can be
introduced \cite{Bern:1990ux,Bern:1994zx}. One loop amplitude can be
decomposed into partial amplitudes with single trace
$N_c\mbox{Tr}(X)$ structure and double trace
$\mbox{Tr}(X)\mbox{Tr}(Y)$ structure, while two loop amplitude can
be decomposed into partial amplitudes with leading-color single trace
$N_c^2\mbox{Tr}(X)$, subleading-color single trace $\mbox{Tr}(X)$, double
trace $N_c\mbox{Tr}(X)\mbox{Tr}(Y)$ and triple trace
$\mbox{Tr}(X)\mbox{Tr}(Y)\mbox{Tr}(Z)$ structures. For general
$L$-loop amplitude, there are at most $(L+1)$ traces appearing with
planar or non-planar structures \cite{Bern:1990ux}.

As the case of tree-level, not all  loop partial  amplitudes are
independent to each other.  A traditional method to analyze their
relation is to use $U(1)$-decoupling equations based on the
observation that photon will decouple from theory where there is no
matter field \cite{Bern:1990ux}. The steps to get these relations
are as follows. First we take one
 generator to be $U(1)$, then color traces will reorganize themselves.
Because the full amplitude vanishes and the reorganized color traces
are independent to each other, the corresponding coefficient of each
reduced color trace will be zero too.  Thus we obtain a series of
equations among partial amplitudes, which are called  {\sl
$U(1)$-decoupling equations}. By solving these equations, we could
express some partial amplitudes by other partial amplitudes.
However, we know that there are  relations  that can not be
discovered by $U(1)$-decoupling method as given in
\cite{Bern:1994zx} for one-loop partial amplitudes, where string
inspired arguments have to  be used. Are there other useful methods, besides $U(1)$-decoupling method, for loop partial amplitudes?

To answer the question, first we notice that loop-level partial
amplitudes can also be studied by direct calculations using, for
example, Feynman diagrams or other methods. Among these methods,
unitarity cut method
\cite{Landau:1959fi,Bern:1994cg,Bern:1994zx,Britto:2005ha,Anastasiou:2006jv}
and generalized unitarity cut method
\cite{Britto:2004nc,Buchbinder:2005wp,Cachazo:2008vp,Cachazo:2008hp,ArkaniHamed:2009dn}
(i.e, the leading singularity method) have been proved to be
particularly useful to obtain loop results (especially one-loop
results) from tree-level input\footnote{BCJ-like relation in loop
level has also been investigated in several works, see
\cite{BjerrumBohr:2011xe,Bern:2010ue,Bern:2010yg,Bern:2011ia,Carrasco:2011hw}.}.

Encouraged by the success of unitarity cut method in the calculation
of one-loop scattering amplitudes,  we find that  just like BCFW
recursion relation has been used to prove relations between
tree-level partial amplitudes,  unitarity cut method can also been
used to understand relations between loop-level partial amplitudes.
In other words, besides the $U(1)$-decoupling method, there is
indeed another method available to our investigation.

The skeleton of this paper is as follows. In section two we study
one-loop amplitude from both $U(1)$-decoupling and unitarity cut
method. Especially we have reproved the KK-like relation
(\ref{sublanswer}) for one-loop partial amplitudes using the
unitarity cut method.  In section three, two-loop amplitude is
investigated similarly, where possible KK relation for subleading-color
single trace partial amplitudes is discussed for six, seven and
eight points. In the last section, a summary is given, as well as
some comments on possible future directions. Some calculation
details and checking are given in two Appendixes.

{\bf Notation:} For simplicity, in this paper  we will consider the
$U(N)$ gauge group instead of $SU(N)$ gauge group. The $U(N)$
generators are a set of hermitian $N\times N$ matrices with
normalization $\mbox{Tr}(T^aT^b)=\delta^{ab}$, and the structure
constant is defined as
\bea [T^a,T^b]=if^{abc}T^c~.~~~\eea
Thus the Fierz identities of $U(N)$ gauge theory are
\bea \sum_{a}\mbox{Tr}(T^aX)\mbox{Tr}(T^aY)=\mbox{Tr}(XY)~,~~~
\sum_a\mbox{Tr}(T^aXT^aY)=\mbox{Tr}(X)\mbox{Tr}(Y)~,~~~\label{Trace-sum-1}\eea
where one special case is
\bea \sum_a
\mbox{Tr}(XT^aT^aY)=N_c\mbox{Tr}(XY)~.~~~\label{Trace-sum-2}\eea
These relations are useful when we discuss the color structure using
the unitarity cut method.

%%%%%%%%%%%%%%%%%%
\section{Partial amplitudes of one loop amplitude}
%%%%%%%%%%%%%%%%%%%%%

The color decomposition of loop amplitude in $U(N)$ gauge theory can
be understood from view of $U(N)$ open string, whose
infinite-tension limit reduces to gauge theory \cite{Bern:1990ux}.
One can also sketch the various trace structures of color
decomposition from arguments based on Feynman diagram analysis.
Different from tree amplitudes, double trace structure appears in
one-loop level, and the corresponding partial amplitudes can be expressed as linear combination of primitive
(partial) amplitudes, i.e., amplitudes of single trace structure.
Schematically, the color decomposition of $n$-point one-loop
amplitude for $U(N)$ gauge theory can be written as
\cite{Bern:1994zx}
\bea {\cal A}_n^{full}\left( \{k_i,\lambda_i,a_i\}\right) =
  \sum_{J} n_J\,\sum_{m=0}^{\lfloor{n/2}\rfloor}
      \sum_{\sigma \in S_n/S_{n;m}}
     \mbox{Gr}_{n-m,m}\left( \sigma \right)\,A_{n-m,m}^{[J]}(\sigma_1,\sigma_2,
     \ldots,\sigma_{n-m};\sigma_{n-m+1},\ldots,\sigma_n)~,
~~~\label{one-loop-color-decomposition} \eea
where ${\lfloor{x}\rfloor}$ is the largest integer less than or
equal to $x$ and $n_J$ is the number of particles of spin $J$. The
color structure for primitive amplitude is (For convenience we
abbreviate $\mbox{Tr}\left( T^{a_1}\cdots T^{a_n}\right)$ as
$\mbox{Tr}\left( a_1,\cdots,a_n\right)$)
\bea \mbox{Gr}_{n,0} = N_c\mbox{Tr}\left( a_1,\cdots
,a_n\right)~,~~~\nonumber\eea
and for other partial amplitudes is
\bea \mbox{Gr}_{n-m,m} = \mbox{Tr}\left(a_1,\cdots,n-m\right)\,
\mbox{Tr}\left( n-m+1,\cdots ,n\right)~.~~~\nonumber\eea
$S_n$ is the set of all permutations of $n$ objects, and $S_{n;m}$
is the subset leaving $\mbox{Gr}_{n-m,m}$ invariant.  If the gauge
group is $SU(N)$, then there is no $\mbox{Gr}_{n-1,1}$ term since
$\mbox{Tr}(T_a)=0$. However, the partial amplitude $A_{n-1,1}$ is
well defined and non-zero. To make $A_{n-1,1}$ explicit in the
expression, we consider $U(N)$ gauge theory instead of $SU(N)$.

It is found that  the partial amplitudes $A_{n-m,m}$ of double trace
structure $\mbox{Gr}_{n-m,m}(m\neq 0)$ have algebraic relation with
primitive amplitudes $A_{n,0}$ of single trace, i.e., $A_{n-m,m}$
can be expressed as linear combination of $A_{n,0}$, thus the
computing of primitive amplitudes is enough for constructing full
one-loop amplitude. The relation is given by \cite{Bern:1994zx}
\bea
 A_{n-m,m}(\alpha_1,\alpha_2,\ldots,\alpha_{n-m};\beta_{1},\ldots,\beta_m)\ =\
 (-1)^{m} \sum_{\sigma\in COP\{\alpha\}\bigcup \{\beta^T\}} A_{n,0}(\sigma)
~,~~~\label{sublanswer} \eea
where $\beta^T$ is the set of $\beta$ with reversed ordering, and
$COP\{\alpha\}\bigcup \{\beta^T\}$ is the set of all permutations
of $\{\alpha,\beta^T\}$ preserving the cyclic ordering inside the
set $\a$ and $\b^T$, but allowing all possible relative orderings
between two sets $\alpha$ and $\b^T$.
%It is also obvious that the
%permutation $\sigma$ is defined up to cyclic symmetry, thus to fix
%the freedom, we can set, for example, one element of $\beta$ at the
%first position.
This  equation can be expected from analysis of open string
amplitude \cite{Bern:1990ux} or  from the view of new
color-decomposition discussed in \cite{DelDuca:1999rs}. The aim of
this section is to understand (\ref{sublanswer}) using unitarity cut
method.

Before going to details, let us give some remarks. As shown in
\cite{DelDuca:1999rs}, the (\ref{sublanswer}) is the direct
consequence of color Jacobi structure  at one-loop level. 
The Jaboci structure is easily seen from $f^{abc}$ but not so
transparent for color-ordered partial amplitudes at tree and loop
levels. For example, Jacobi structure of tree-level color-ordered
amplitude is hidden implicitly in the KK relation and BCJ relation.
Our discussions in this section will provide another point of view
to understand the same question, although our method  is a little bit
circuitous.

%%%%%%%%%%%%%%%%%%%%%%%%
\subsection{Revisit of $U(1)$-decoupling method}
%%%%%%%%%%%%%%%%%%%%%%%

Before discussing the unitarity cut method, let us revisit the
$U(1)$-decoupling equation carefully. We will show that with the
$U(1)$-decoupling equation, a relation such like (\ref{sublanswer})
will not emerge. Thus new thought is needed to understand
(\ref{sublanswer}).

%%%%%%%%%%%%%%%%%%%%%%
\subsubsection{The general $U(1)$-decoupling equations}
%%%%%%%%%%%%%%%%%%%%%

The central idea of $U(1)$-decoupling equation is to choose one of
generators to be $U(1)$, then the original trace structure
(\ref{one-loop-color-decomposition}) will reorganize itself to new
color trace structure. Since photon does not interact with others,
coefficients of new color traces will be zero.

To demonstrate, let us consider four-point amplitude which has three
kinds of partial amplitudes: one  with single trace structure
$A_{4,0}$ and the other two  with double trace structure, $A_{3,1}$ and
$A_{2,2}$. Their corresponding color structures can be abbreviated
as $(4|0),(3|1)$ and $(2|2)$.  By setting one generator as identity,
these color structures reduce to $(3|0)$ and $(2|1)$ as follows
\bea (4|0)\to(3|0)&~,~~~&(3|1)\to(2|1) ~\mathrm{or}
~(3|0)~,~~~(2|2)\to(2|1)~.~~~\eea
More explicitly,  with $T_4$ being $U(1)$,  the reduced
color structure $(3|0)=N_c\mbox{Tr}(1,2,3)$ gives the following
$U(1)$-decoupling equation
\bea
A_{4,0}(1,2,3,4)+A_{4,0}(1,2,4,3)+A_{4,0}(1,4,2,3)+A_{3,1}(1,2,3;4)=0~,~~~
\label{U1-4-30}\eea
while the reduced color structure $(2|1)=\mbox{Tr}(1,2)\mbox{Tr}(3)$
gives (there are other $(2|1)$ trace structures)
\bea A_{3,1}(1,2,4; 3)+ A_{3,1}(1,4,2;3)+
A_{2,2}(1,2;3,4)=0~.~~~\label{U1-4-21}\eea
Using the equation (\ref{U1-4-30}) we can solve $A_{3,1}$ as a linear
combination of $A_{4,0}$. Then together with (\ref{U1-4-21}), we can finally
solve $A_{2,2}$ as a linear combination of $A_{4,0}$.

For general $n$, with  trace structure  (where we have put trace
structure implicitly  as parameters of partial amplitudes $A$)
\bea {\cal A}^{1-loop} & = & \sum N_c A_{n,0}(\sigma_1,...,\sigma_n)
+\sum_{m} A_{n-m,m}(\sigma_1,...,\sigma_{n-m};
\beta_1,...,\beta_{n})~,~~~ \eea
when we take, for example, $T^n$ to be $U(1)$, then $(n-m|m)$
structure will reduce to either $(n-m|m-1)$ or $(n-m-1|m)$
structures depending on where $T^n$ locates. Collecting all terms
having the same reduced color structure we get one $U(1)$-decoupling
equation. By taking different reduced color structures and different
$T^a$ to be $U(1)$ we can get different equations.

We can go further by taking more than one $T^a$ to be $U(1)$.
However, since after one $U(1)$ reduction,  all coefficients of
reduced color structures are zero already, taking more than one
generator to be  $U(1)$ does not give new relations. Thus to get all
independent $U(1)$-decoupling equations, we just need to take one
$T^a$ to be $U(1)$ with $a=1,...,n$ and write down all equations
obtained by this way.

Having above general discussions, now let us write down equations
obtained by taking $T^n$ to be $U(1)$. The reduced trace structure
$\mbox{Tr}(1,\ldots,m-1)\mbox{Tr}(m,\ldots,n-1)$  will  receive
contributions from partial amplitudes of original trace structures
$(n-m|m)$ and $(n-m+1|m-1)$, so the corresponding $U(1)$-decoupling
equation is
\bea 0=\sum_{\sigma\in
~cyclic}A_{n-m,m}(\sigma_1,\ldots,\sigma_{m-1},n;m,\ldots,n-1)+\sum_{\sigma\in
~cyclic}A_{n-m+1,m-1}(1,\ldots,m-1;\sigma_m,\ldots,\sigma_{n-1},n)~,~~~\label{U1-n-gen}\eea
where $1\leq m\leq {\lfloor{n/2}\rfloor}$. When  $m=1$,
(\ref{U1-n-gen}) can be used to solve  $A_{n-1,1}$ by  single trace
part $A_{n,0}$ as
\bea A_{n-1,1}(1,\ldots,n-1;n)=-\sum_{\sigma\in
~cyclic}A_{n,0}(\sigma_1,\ldots,\sigma_{n-1},n)~.~~~\eea
When  $m=2$, (\ref{U1-n-gen}) contains only one $A_{n-2,2}$ thus we
can solve it as
\bea
A_{n-2,2}(1,\ldots,n-2;n-1,n)&=&-\sum_{\sigma\in~cyclic}A_{n-1,1}(\sigma_1,\ldots,\sigma_{n-2},n;n-1)\nonumber\\
&=&\sum_{\sigma\in~cyclic}\sum_{\alpha\in~cyclic}A_{n,0}(\alpha_{\sigma_1},\ldots,\alpha_{\sigma_{n-2}},\alpha_{n},n-1)\nonumber\\
&=&\sum_{\sigma\in COP\{1,\ldots,n-2\}\cup
\{n,n-1\}}A_{n,0}(\sigma)~.~~~\eea
Things become more complicated when $m\geq 3$. The reason is that
amplitudes $A_{n-m,m}$ always appear in group in (\ref{U1-n-gen})
and there is no way to separate them. As we will see explicitly in
six-point example, $U(1)$-decoupling equations are not enough to
solve $A_{n-m,m}$ by $A_{n,0}$ as given in (\ref{sublanswer}), but
they do give hints.

%%%%%%%%%%%%%%%%%
\subsubsection{Analysis of six-point amplitude}
%%%%%%%%%%%%%%%%%%

As we have mentioned, when $m\geq 3$, it is impossible to solve all
$A_{n-m,m}$ by $A_{n,0}$ directly through  $U(1)$-decoupling
equations. To see it clearly, we take the simplest example where the
phenomenon happens, i.e., the one-loop six-point amplitude. First we
write down $U(1)$-decoupling relation for $n=6$ explicitly
\cite{Bern:1990ux} as
\bea 0&= & \left[A_{5,1} (\sigma_1,...,\sigma_5;6)+\sum_{cyclic}
A_{6,0}(6,\sigma_1,...,\sigma_5) \right]~,~~~\label{n=6-1}\\
0 & = &
\sum_{cyclic~\sigma}A_{5,1}(6,\sigma_1,...,\sigma_4;\beta_1)+
A_{4,2}(\sigma_1,...,\sigma_4;6,\beta_1)~,~~~\label{n=6-2}\\
0 & = & \sum_{cyclic~\sigma} A_{4,2}(6,\sigma_1,...,\sigma_3;
\beta_1,\beta_2)+ \sum_{cyclic~\beta}A_{3,3} (\sigma_1,...,\sigma_3;
6,\beta_1,\beta_2)~,~~~\label{n=6-3}\eea
where $T^6$ has been set as $U(1)$ identity. Using (\ref{n=6-1}) we
can solve any $A_{5,1}$ by  partial amplitudes $A_{6,0}$:
\bea A_{5,1} (1,2,3,4,5;6) & = &
-A_{6}(6,1,2,3,4,5)-A_{6}(6,5,1,2,3,4)-A_{6}(6,4,5,1,2,3) \nn & &
-A_{6}(6,3,4,5,1,2)-A_{6}(6,2,3,4,5,1)~.~~~\label{A51}\eea
Having the $A_{5,1}$ we can use (\ref{n=6-2}) to solve
$A_{4,2}$ by partial amplitudes $A_{6,0}$ as
\bea A_{4,2}(1,2,3,4;5,6)=\sum_{\sigma\in COP\{1,2,3,4\}\cup
\{6,5\}}A_{6,0}(\sigma)~.~~~\eea
%
%Then the consistence can be easily understood from the cyclic
%ordered permutations.

The tricky part is $A_{3,3}$. From equation (\ref{n=6-3}) we
have
\bea A_{3,3}(1,2,3; 6,4,5)+ A_{3,3}(1,2,3; 6,5,4)& = &
X_1=-\sum_{\sigma\in~cyclic}A_{4,2}(6,\sigma_1,\sigma_2,\sigma_3;4,5)~,~~~\label{A33-1}\\
A_{3,3}(1,3,2; 6,4,5)+ A_{3,3}(1,3,2; 6,5,4) &= &
X_2=-\sum_{\sigma\in~cyclic}A_{4,2}(6,\sigma_1,\sigma_3,\sigma_2;4,5)~,~~~\label{A33-2}\eea
where $X_1, X_2$ are, in fact, a linear combination of $A_{6,0}$.
Taking leg $5$ or $4$ to be photon we can obtain other two
equations, which are similar to (\ref{A33-1}) and (\ref{A33-2}) by
relabeling  $\{4,5,6\}\to \{4,6,5\}$ and $\{4,5,6\}\to \{6,4,5\}$.
However, the left hand sides of (\ref{A33-1}) and (\ref{A33-2}) are
invariant directly under above relabeling, while the invariant of the right hand
sides can be seen explicitly only  when expanded as a linear combination of
$A_{6,0}$.
%But since the amplitudes are cyclic permutation invariant
%within each trace, the above relabeling gives the same expression in
%the left hand side of (\ref{A33-1}) and (\ref{A33-2}). The
%equivalence of right hand side can be checked directly by expanding
%them into $A_{6,0}$.
Thus there are no new relations coming from taking $T^4, T^5$ to
be $U(1)$.

Using  $T_1$ to be  $U(1)$ we get another equations
\bea A_{3,3}(1,2,3;4,5,6)+A_{3,3}(1,3,2;4,5,6)&= &\W
X_1=-\sum_{\sigma\in~cyclic}A_{4,2}(1,\sigma_4,\sigma_5,\sigma_6;2,3)~,~~~\label{A33-7}\\
A_{3,3}(1,2,3;4,6,5)+A_{3,3}(1,3,2;4,6,5)&= &\W
X_2=-\sum_{\sigma\in~cyclic}A_{4,2}(1,\sigma_4,\sigma_6,\sigma_5;2,3)~.~~~\label{A33-8}
\eea
Similarly it can be shown that taking  $T^2,T^3$ to be $U(1)$ will
not give new relations.

Equations (\ref{A33-1}), (\ref{A33-2}), (\ref{A33-7})  and
(\ref{A33-8}) are all independent ones we can obtain, provided that the legs $1,2,3$ in $A_{3,3}$ are all in one trace and the others are all in another trace.
% from
%$U(1)$-decoupling equations involving $A_{3,3}$ with leg $1,2,3$ in
%one trace and $4,5,6$ in another trace.
 There are four partial
amplitudes $t_1=A_{3,3}(1,2,3;4,5,6)$, $t_2=A_{3,3}(1,3,2;4,5,6)$,
$t_3=A_{3,3}(1,2,3;4,6,5)$, $t_4=A_{3,3}(1,3,2;4,6,5)$, and four
equations, which, when written in matrix form, are
\bea \left( \begin{array}{c} X_1 \\ X_2 \\ \W X_1 \\ \W X_2
\end{array}\right)=\left( \begin{array}{cccc} 1 & 0 & 1 & 0 \\ 0 & 1
& 0 & 1\\ 1 & 1 & 0 & 0 \\ 0 & 0 & 1 & 1
\end{array}\right)\left( \begin{array}{c} t_1 \\ t_2 \\ t_3 \\ t_4
\end{array}\right)~.~~~\eea
This one has unique solution when and only when the determinant is
non-zero. However, it is easy to check that the determinant is
indeed zero and we find following solution:
\bea t_1= t_4+\W X_1-X_2,~~~t_2=-t_4+X_2,~~~t_3=-t_4-\W
X_1+X_1+X_2~,~~~\eea
which indicates the impossibility of expressing all partial
amplitudes $A_{3,3}$ as a linear combination of $A_{6,0}$ from
$U(1)$-decoupling equations\footnote{The consistent condition
requires $\W X_1+\W X_2-X_1-X_2=0$, which can be verified trivially
by writing $X_i, \W X_i$ as sum of $A_{6,0}$.}.

Although $U(1)$-decoupling equations can not solve $A_{3,3}$ as
combinations of $A_{6,0}$, they may give some hints to the solution
(\ref{sublanswer}). The key is to write the $X_i, \W X_i$ as
\bea X_1=(456)\bigcup (123)+(465)\bigcup(123),~~~~X_2= (456)\bigcup
(132)+(465)\bigcup(132) ~,~~~\label{X1} \eea
and
\bea \W X_1= (123)\bigcup (456)+(132)\bigcup (456),~~~~\W
X_2=(123)\bigcup (465)+(132)\bigcup (465)~,~~~\eea
where we have simplified $\sum_{\sigma \in COP \{1,2,3\}\bigcup
\{4,5,6\}} A(\sigma)$ as $(456)\bigcup (123)$. With these rewriting,
it is very natural to make some identifications. From (\ref{A33-1})
and (\ref{A33-2}) there are two choices can be made. The choice (A)
is given by
\bea A_{3,3}(1,2,3;4,5,6)& = &  (1,2,3)\bigcup
(4,5,6),~~~~A_{3,3}(1,2,3;6,5,4)= (1,2,3)\bigcup (6,5,4)~,~~~\nn
A_{3,3}(1,3,2;4,5,6)& = &  (1,3,2)\bigcup
(4,5,6),~~~~A_{3,3}(1,3,2;6,5,4)= (1,3,2)\bigcup
(6,5,4)~,~~~\label{Choice-A}\eea
while the choice (B) is given by
\bea A_{3,3}(1,2,3;4,5,6)& = &  (1,2,3)\bigcup
(6,5,4),~~~~A_{3,3}(1,2,3;6,5,4)= (1,2,3)\bigcup (4,5,6)~,~~~\nn
A_{3,3}(1,3,2;4,5,6)& = &  (1,3,2)\bigcup
(6,5,4),~~~~A_{3,3}(1,3,2;6,5,4)= (1,3,2)\bigcup
(4,5,6)~.~~\label{Choice-B}\eea
Both choices are consistent with (\ref{A33-7}) and (\ref{A33-8}) if
we notice that the color-order reversed relation means $(\a \bigcup
\b)= (-)^n ( \a^T \bigcup \b^T)$, i.e., $(1,2,3)\bigcup(4,5,6)=
(1,3,2)\bigcup (6,5,4)$. However, the right solution is the choice
(B). Our six-point example may be too special and when we move to
higher points,  hints will be more explicit.

%%%%%%%%%%%%%%%%%%%%%%%%%
\subsection{Unitarity cut method}
%%%%%%%%%%%%%%%%%%%%%%%%%%

Unitarity cut method has been proved to be very useful for
calculations of loop amplitudes. For one-loop amplitude,
Passarino-Veltman reduction shows that any one-loop amplitude can be
expanded to some known basis with rational coefficients \cite{PV}.
The basis contains scalar integrals with topologies as tadpole,
bubble, triangle and box in pure 4D theory (in this case we need to
add rational remaining functions) or in general
$(4-2\epsilon)$-dimension with pentagon\footnote{There are some
recent works on the basis of two-loop amplitudes, for example see
\cite{Gluza:2010ws}. The basis for general multiloop amplitudes is
not clear yet, but we can still get useful information for analyzing
by using unitarity cut method. }. Thus  loop calculations are
reduced to finding coefficients of corresponding basis
\cite{Bern:1994cg}. The advantage of unitarity cut method for
calculations is that inputs are on-shell tree-level amplitudes which
have all desired symmetries, including  gauge symmetry.

Much calculations that have been done using unitarity cut method are
for color-ordered partial amplitudes. However, as we will show in
this subsection, unitarity cut method is also very useful for
calculations of whole amplitudes as well as the understanding of
color structure of general loop amplitudes.

To start, let us write down the expression for calculating the whole
one-loop amplitude at a given cut channel with momentum
$K$\footnote{Similar idea has been discussed in paper
\cite{Luo:2004ss, Luo:2004nw}, where the unitarity cut method plus
the CSW rule has been applied to the one-loop MHV amplitudes.}

\bea
\mathcal{A}_{n}^{1-loop}|_{cut}=\sum_{states~of~\ell_1,\ell_2}{\cal
A}_{L}^{full~tree}(\ell_1,\alpha_L,\ell_2,\beta_L){\cal
A}_{R}^{full~tree}(-\ell_1,\alpha_R,-\ell_2,\beta_R)~,~~~\label{one-loop-cut}\eea
where the ordering of $\{\ell_1,\alpha_L,\ell_2,\beta_L\}$ does not
mean anything because the input is the full on-shell tree amplitude
${\cal A}$. To uniquely determine the full one-loop amplitude, we
need to calculate  all possible cut channels in unitarity cut
method. A few of remarks are in order before we go on. First we have
used double cuts,
 which can not access
 tadpole coefficients. Fortunately, for gauge theory which is
massless, there is no tadpole contribution. Secondly, we require
that there should be at least two external legs in $A_L$ or $A_R$
for each cut channel, which is satisfied for massless particles.
Thirdly, to get complete result for one-loop amplitude using
unitarity cut method, the cut momenta $\ell_1, \ell_2$ should be in
general $(4-2\epsilon)$-dimension. Thus we have assumed  properties
which we will use later, such as the KK relation, will hold in
general D-dimension. This is true for gauge theory because as we
have remarked, KK relation is a pure group theory relation.
%If KK
%relation holds only for 4D, our results will be true only up to
%rational remaining terms. For pure gauge theory we know there are
%nontrivial remaining terms, but the ${\cal N}=4$ theory does not
%have.

With above clarifications,  we will discuss how various trace structures
of one-loop amplitude, i.e., the single trace and double trace
structures, show up in the unitarity cut method. Substituting two full
tree amplitudes with their color-decompositions in
(\ref{one-loop-cut}), and noticing that the sum over states of
$\ell_1,\ell_2$ include the sum over colors, we have (where for
simplicity we have  written $\ell_1$ for $T^{\ell_1}$)
\bea
\sum_{\ell_2}\sum_{\ell_1}\mbox{Tr}(\ell_1,\alpha_L,\ell_2,\beta_L)
\mbox{Tr}(\ell_1,\alpha_R,\ell_2,\beta_R)=\sum_{\ell_2}
\mbox{Tr}(\alpha_L,\ell_2,\beta_L,\alpha_R,\ell_2,\beta_R)=
\mbox{Tr}(\beta_L,\alpha_R)\mbox{Tr}(\b_R,\a_L)~,~~~\label{one-loop-uni-trace}\eea
where (\ref{Trace-sum-1}) has been used. Equation
(\ref{one-loop-uni-trace}) is the general double trace color
structure and when we
 set $\{\alpha_L,\beta_R\}$ or $\{\beta_L,\alpha_R\}$ 
empty, it is reduced to the single trace structure
$N_c\mbox{Tr}(\beta_L,\alpha_R)$ or
$N_c\mbox{Tr}(\alpha_L,\beta_R)$. Correspondingly, coefficients for
the double trace structure
$\mbox{Tr}(\alpha_L,\beta_R)\mbox{Tr}(\beta_L,\alpha_R)$ will have
contributions from following cut-input
\bea \sum_{h_1, h_2}A_L(\ell_1^{h_1},\alpha_L,\ell_2^{h_2},\beta_L)
A_R(-\ell_1^{-h_1},\alpha_R,-\ell_2^{-h_2},\beta_R)~,~~\label{One-loop-A-form}\eea
where $A_L, A_R$ are color-ordered tree-level partial amplitudes.
For the same trace structure, another cut-input,
\bea \sum_{h_1, h_2}A_L(\ell_1^{h_1},\b_L,\ell_2^{h_2},\a_L)
A_R(-\ell_1^{-h_1},\b_R,-\ell_2^{-h_2},\a_R)~,~~\label{One-loop-A-form-2}\eea
also gives contribution,
which comes from exchanging set $\a$ and $\b$, or
equivalently, exchanging $\ell_1, \ell_2$. In principle we
should sum up these two contributions together. However, for
one-loop, it can be seen that above two inputs
(\ref{One-loop-A-form}) and (\ref{One-loop-A-form-2}) give the same
contributions. Using this freedom we can write (\ref{one-loop-cut})
formally as
\bea
\mathcal{A}_n^{1-loop}|_{cut}&=&\sum_{L,R}\left[\sum_{P\{\alpha_{L}(1),
\ell_2,\beta_L\}}\mbox{Tr}(\ell_1,\alpha_L(1),\ell_2,\beta_L)A_L(\ell_1,\alpha_L(1),
\ell_2,\beta_L)\right.~~~~\label{np1loop-2}\\
&&~~~~~~~~~~~\times\left.\sum_{P\{\alpha_{R},\ell_2,\beta_R\}}\mbox{Tr}
(\ell_1,\alpha_R,\ell_2,\beta_R)A_R(-\ell_1,\alpha_R,-\ell_2,\beta_R)\right]\nonumber\\
&&+\{\ell_1\leftrightarrow\ell_2\}~,~~~\nonumber\eea
where $\a_L(1)$ means that particle $1$ belongs to set $\a_L$ and the
sum $\sum_{L,R}$ means we need to consider all allowed double cuts. The whole amplitude $\mathcal{A}_n^{1-loop}$ is not equal to the right hand side of (\ref{np1loop-2}) but is determined by the right hand side through the unitarity cut method.
%The
%equation (\ref{np1loop-2}) is not really an identity between left
%and right hand sides, and we just mean that left hand side is
%determined by right hand side using the unitarity cut method and the
%sum $\sum_{L,R}$ means we need to consider all allowed double cuts.
Also as we have remarked, since all  calculations are the same after
exchanging $\ell_1,\ell_2$ we could   only consider the first term of
(\ref{np1loop-2}).

Tree-level partial amplitudes in (\ref{np1loop-2}) are with all
possible color-orderings, but by using  $U(1)$-decoupling relation,
KK relation and BCJ relation among them, we may establish relations
between different trace structures. More explicitly, assuming there
is  a combination of one-loop partial amplitudes of different trace
structures
\bea I & = & \sum_i c_i A_{n-i,i}~, \eea
if under  all possible double cuts,  above combination is zero, i.e., $\sum_i c_i
A_{L,i} A_{R,i}=0$,  then we can claim that  the combination $I$ is
zero, i.e., there is a nontrivial relation among these one-loop
partial amplitudes, up to three points discussed below equation
(\ref{one-loop-cut}).

With these general discussions, now we will apply the unitarity cut
method to investigate  relations among one-loop partial amplitudes.
We will study the four-point amplitude in detail as an example and
then  give a proof of result (\ref{sublanswer}).

%%%%%%%%%%%%%%%%%%%%%
\subsubsection{The example of  four-point amplitude}
%%%%%%%%%%%%%%%%%%%%%%%%

As an illustration of unitarity cut method, let us give a detail
analysis of four-point one-loop amplitude. After fixing leg $1$ in
the left tree amplitudes (we can always do that), and decomposing the
full tree amplitude as partial amplitudes,
 (\ref{one-loop-cut})  becomes
\bea \mathcal{A}_4^{1-loop}|_{cut}&=&~~\sum_{P\{\ell_2,1,2\}}
\mbox{Tr}(\ell_1,\ell_2,1,2)A_{L}(\ell_1,\ell_2,1,2)\sum_{P\{-\ell_2,3,4\}}
\mbox{Tr}(\ell_1,\ell_2,3,4)A_{R}(-\ell_1,-\ell_2,3,4)\nonumber\\
&&+\sum_{P\{\ell_2,1,3\}}
\mbox{Tr}(\ell_1,\ell_2,1,3)A_{L}(\ell_1,\ell_2,1,3)\sum_{P\{-\ell_2,2,4\}}
\mbox{Tr}(\ell_1,\ell_2,2,4)A_{R}(-\ell_1,-\ell_2,2,4)\nonumber\\
&&+\sum_{P\{\ell_2,1,4\}}
\mbox{Tr}(\ell_1,\ell_2,1,4)A_{L}(\ell_1,\ell_2,1,4)\sum_{P\{-\ell_2,2,3\}}
\mbox{Tr}(\ell_1,\ell_2,2,3)A_{R}(-\ell_1,-\ell_2,2,3)\label{4p1loop}~,~~~\eea
where using cyclic symmetry we can always fix $\ell_1$ at the first
position and $P\{\alpha\}$ means all  permutations on set
$\{\alpha\}$. There are totally $3\times 3!\times 3!=108$ terms in
the right hand side of (\ref{4p1loop}), and they contribute to
different trace structures. These terms can be written down in the
form (\ref{np1loop-2})  as
\bea \mathcal{A}_4^{1-loop}|_{cut}
&=&\Big(\mbox{Tr}(\ell_1,1,\ell_2,2)A_L(\ell_1,1,\ell_2,2)+\mbox{Tr}(\ell_1,1,2,\ell_2)A_L(\ell_1,1,2,\ell_2)~~~~\label{4p1loopv2}\\
&&+\mbox{Tr}(\ell_1,2,1,\ell_2)A_L(\ell_1,2,1,\ell_2)\Big)\times\sum_{P\{-\ell_2,3,4\}}\mbox{Tr}(\ell_1,\ell_2,3,4)A_R(-\ell_1,-\ell_2,3,4)\nonumber\\
&&+\Big(\mbox{Tr}(\ell_1,1,\ell_2,3)A_L(\ell_1,1,\ell_2,3)+\mbox{Tr}(\ell_1,1,3,\ell_2)A_L(\ell_1,1,3,\ell_2)\nonumber\\
&&+\mbox{Tr}(\ell_1,3,1,\ell_2)A_L(\ell_1,3,1,\ell_2)\Big)\times\sum_{P\{-\ell_2,2,4\}}\mbox{Tr}(\ell_1,\ell_2,2,4)A_R(-\ell_1,-\ell_2,2,4)\nonumber\\
&&\Big(\mbox{Tr}(\ell_1,1,\ell_2,4)A_L(\ell_1,1,\ell_2,4)+\mbox{Tr}(\ell_1,1,4,\ell_2)A_L(\ell_1,1,4,\ell_2)\nonumber\\
&&+\mbox{Tr}(\ell_1,4,1,\ell_2)A_L(\ell_1,4,1,\ell_2)\Big)\times\sum_{P\{-\ell_2,2,3\}}\mbox{Tr}(\ell_1,\ell_2,2,3)A_R(-\ell_1,-\ell_2,2,3)\nonumber\\
&+&\{\ell_1\leftrightarrow\ell_2\}\nonumber~.~~~\eea
Now let us consider contributions to various trace structures using
(\ref{4p1loopv2}).

Firstly we consider the single trace structure. For example, the
single trace $N_c\mbox{Tr}(1,2,3,4)$ comes from four terms:
\bea
\mbox{Tr}(\ell_1,1,2,\ell_2)\mbox{Tr}(\ell_1,\ell_2,3,4)A_L(\ell_1,1,2,\ell_2)A_R(-\ell_1,-\ell_2,3,4)~,~~~\nonumber\\
\mbox{Tr}(\ell_1,4,1,\ell_2)\mbox{Tr}(\ell_1,\ell_2,2,3)A_L(\ell_1,4,1,\ell_2)A_R(-\ell_1,-\ell_2,2,3)~,~~~\nonumber\eea
and other two terms with $\{\ell_1\leftrightarrow\ell_2\}$. Thus we
can write 
\bea A_{4,0}(1,2,3,4)|_{cut}&=&
A_L(\ell_1,1,2,\ell_2)A_R(-\ell_1,-\ell_2,3,4)+A_L(\ell_1,4,1,\ell_2)A_R(-\ell_1,-\ell_2,2,3)+\{\ell_1\leftrightarrow\ell_2\}~.~~~\label{4pan1}\eea
For the color ordering
$(1,2,3,4)$ we can determine the amplitude by considering two double
cuts: cut $K_{12}$ and cut $K_{41}$, and they are exactly the two
terms we have written down in (\ref{4pan1}).

Next we consider double trace structure. For $A_{3,1}(1,2,3;4)$
with trace structure $\mbox{Tr}(1,2,3)\mbox{Tr}(4)$,  terms that
contribute to this trace structure are
\bea &&\mbox{Tr}(\ell_1,1,2,\ell_2)\mbox{Tr}(\ell_1,4,\ell_2,3)A_L(\ell_1,1,2,\ell_2)A_R(-\ell_1,4,-\ell_2,3)~,~~~\nonumber\\
&&\mbox{Tr}(\ell_1,3,1,\ell_2)\mbox{Tr}(\ell_1,4,\ell_2,2)A_L(\ell_1,3,1,\ell_2)A_R(-\ell_1,4,-\ell_2,2)~,~~~\nonumber\\
&&\mbox{Tr}(\ell_1,1,\ell_2,4)\mbox{Tr}(\ell_1,\ell_2,2,3)A_L(\ell_1,1,\ell_2,4)A_R(-\ell_1,-\ell_2,2,3)~,~~~\nonumber\eea
and other three terms with $\{\ell_1\leftrightarrow\ell_2\}$. Thus
we can get
\bea
A_{3,1}(1,2,3;4)|_{cut}&=&A_L(\ell_1,1,2,\ell_2)A_R(-\ell_1,4,-\ell_2,3)+A_L(\ell_1,3,1,\ell_2)A_R(-\ell_1,4,-\ell_2,2)\nonumber\\
&&+A_L(\ell_1,1,\ell_2,4)A_R(-\ell_1,-\ell_2,2,3)+\{\ell_1\leftrightarrow\ell_2\}~.~~~\label{4pan2}\eea
Then we can  use KK relation (\ref{KK}) to put
$\ell_1,\ell_2$ at the two ends
\bea
A(\ell_a,i,\ell_b,j)=-A(\ell_a,i,j,\ell_b)-A(\ell_a,j,i,\ell_b)~,~~~\eea
so (\ref{4pan2}) can be rewritten as
\bea A_{3,1}(1,2,3;4)|_{cut}&=&
-A_L(\ell_1,1,2,\ell_2)A_R(-\ell_1,-\ell_2,3,4)-A_L(\ell_1,4,1,\ell_2)A_R(-\ell_1,-\ell_2,2,3)~~~~\label{4pan2p1}\\
&&-A_L(\ell_1,1,2,\ell_2)A_R(-\ell_1,-\ell_2,4,3)-A_L(\ell_1,3,1,\ell_2)A_R(-\ell_1,-\ell_2,2,4)~~~~\label{4pan2p2}\\
&&-A_L(\ell_1,1,4,\ell_2)A_R(-\ell_1,-\ell_2,2,3)-A_L(\ell_1,3,1,\ell_2)A_R(-\ell_1,-\ell_2,4,2)~~~~\label{4pan2p3}\\
&&+\{\ell_1\leftrightarrow\ell_2\}~. \nonumber\eea
Comparing (\ref{4pan2p1}), (\ref{4pan2p2}), (\ref{4pan2p3}) with
(\ref{4pan1}),  it is clear that each line with its
$\{\ell_1\leftrightarrow\ell_2\}$ terms can be identified as one
primitive amplitude, so we get following identity between $A_{4,0}$
and $A_{3,1}$
\bea
A_{3,1}(1,2,3;4)&=&-A_{4,0}(1,2,3,4)-A_{4,0}(1,2,4,3)-A_{4,0}(1,4,2,3)\nonumber\\
&=&-\sum_{\sigma\in~
cyclic}A_{4,0}(\sigma_1,\sigma_2,\sigma_3,4)~.~~~\label{4Iden-1}\eea

Similar argument can be applied to  another double trace structure
$A_{2,2}(1,2;3,4)$. By working out the trace structures we can
identify $A_{2,2}(1,2;3,4)|_{cut}$ as
\bea A_{2,2}(1,2;3,4)|_{cut}&=&
A_L(\ell_1,1,2,\ell_2)A_R(-\ell_1,3,4,-\ell_2)+A_L(\ell_1,1,2,\ell_2)A_R(-\ell_1,4,3,-\ell_2)\nonumber\\
&&+A_L(\ell_1,1,\ell_2,3)A_R(-\ell_1,4,-\ell_2,2)+A_L(\ell_1,1,\ell_2,4)A_R(-\ell_1,3,-\ell_2,2)\nonumber\\
&&+A_L(\ell_1,2,1,\ell_2)A_R(-\ell_1,3,4,-\ell_2)+A_L(\ell_1,2,1,\ell_2)A_R(-\ell_1,4,3,-\ell_2)\nonumber\\
&&+\{\ell_1\leftrightarrow\ell_2\}~.~~~\eea Using the following KK
relation
\bea
&&A(\ell_a,\ell_b,i,j)=A(\ell_a,j,i,\ell_b)~,~~~A(\ell_a,i,\ell_b,j)=-A(\ell_a,i,j,\ell_b)-A(\ell_a,j,i,\ell_b)~,~~~\nonumber\eea
$A_{2,2}(1,2;3,4)|_{cut}$ can be written as
\bea A_{2,2}(1,2;3,4)|_{cut}&=&
A_L(\ell_1,1,2,\ell_2)A_R(-\ell_1,-\ell_2,4,3)+A_L(\ell_1,3,1,\ell_2)A_R(-\ell_1,-\ell_2,2,4)\nonumber\\
&&+A_L(\ell_1,1,3,\ell_2)A_R(-\ell_1,-\ell_2,2,4)+A_L(\ell_1,4,1,\ell_2)A_R(-\ell_1,-\ell_2,3,2)\nonumber\\
&&+A_L(\ell_1,1,2,\ell_2)A_R(-\ell_1,-\ell_2,3,4)+A_L(\ell_1,4,1,\ell_2)A_R(-\ell_1,-\ell_2,2,3)\nonumber\\
&&+A_L(\ell_1,1,3,\ell_2)A_R(-\ell_1,-\ell_2,4,2)+A_L(\ell_1,2,1,\ell_2)A_R(-\ell_1,-\ell_2,3,4)\nonumber\\
&&+A_L(\ell_1,3,1,\ell_2)A_R(-\ell_1,-\ell_2,4,2)+A_L(\ell_1,1,4,\ell_2)A_R(-\ell_1,-\ell_2,2,3)\nonumber\\
&&+A_L(\ell_1,1,4,\ell_2)A_R(-\ell_1,-\ell_2,3,2)+A_L(\ell_1,2,1,\ell_2)A_R(-\ell_1,-\ell_2,4,3)\nonumber\\
&&+\{\ell_1\leftrightarrow\ell_2\}~.~~~\eea
Each line with its $\{\ell_1\leftrightarrow\ell_2\}$ terms in above
result corresponds to one  of primitive amplitude $A_{4,0}$, and the
whole result is nothing but
\bea
A_{2,2}(1,2;3,4)&=&A_{4,0}(1,2,3,4)+A_{4,0}(1,3,2,4)+A_{4,0}(1,3,4,2)\nonumber\\
&&+A_{4,0}(1,2,4,3)+A_{4,0}(1,4,2,3)+A_{4,0}(1,4,3,2)\nonumber\\
&=&\sum_{\sigma\in COP\{1,2\}\cup \{3,4\}}A_{4,0}(\sigma)~.~~~\eea
%
%where $COP\{1,2\}\cup\{3,4\}$ is the set of all permutations of
%$\{1,2,3,4\}$ with $1$ held fixed that preserve the cyclic ordering
%of set $\{1,2\}$ and $\{3,4\}$, while allowing for all possible
%relative orderings of element in $\{1,2\}$ with respect to element
%in $\{3,4\}$.

Until now we have expressed the partial amplitudes of double trace
structure as a linear combination of primitive amplitudes $A_{4,0}$,
but among these primitive amplitudes, how many are  really
independent basis?  Using the unitarity cut method, it is easy to
see that for arbitrary $n$ we have (we will give a proof later)
\bea A_{n,0}(1,2,3,...,n-1,n)=(-)^n
A_{n,0}(n,n-1,...,2,1)~.~~\label{Oneloop-reverse}\eea
Thus for four-point primitive amplitudes, when accounting the cyclic
symmetry and reflection identity (\ref{Oneloop-reverse}), there are
totally $S_4/(Z_4\times 2)=3$ independent primitive amplitudes. Is
there any further relation among these three amplitudes as in the
case of  tree-level amplitudes? In order to answer this question,
let us investigate  cuts of these three amplitudes:
\bean A_{4,0}(1,2,3,4)|_{cut} &= & A_L(\ell_1, 1,2, \ell_2)
A_R(-\ell_2,3,4,-\ell_1)+
A_L(\ell_1, 4,1, \ell_2) A_R(-\ell_2,2,3,-\ell_1)\nonumber\\
&&+A_L(\ell_1, 2,1, \ell_2) A_R(-\ell_2,4,3,-\ell_1)+ A_L(\ell_1,
1,4, \ell_2) A_R(-\ell_2,3,2,-\ell_1)~,~~~\nn
A_{4,0}(1,3,2,4)|_{cut} &= & A_L(\ell_1, 1,3, \ell_2)
A_R(-\ell_2,2,4,-\ell_1)+
A_L(\ell_1, 4,1, \ell_2) A_R(-\ell_2,3,2,-\ell_1)\nonumber\\
&&+A_L(\ell_1, 3,1, \ell_2) A_R(-\ell_2,4,2,-\ell_1)+ A_L(\ell_1,
1,4, \ell_2) A_R(-\ell_2,2,3,-\ell_1)~,~~~\nn
A_{4,0}(1,3,4,2)|_{cut}&= & A_L(\ell_1, 1,3, \ell_2)
A_R(-\ell_2,4,2,-\ell_1)+
A_L(\ell_1, 2,1, \ell_2) A_R(-\ell_2,3,4,-\ell_1)\nonumber\\
&&+A_L(\ell_1, 3,1, \ell_2) A_R(-\ell_2,2,4,-\ell_1)+ A_L(\ell_1,
1,2, \ell_2) A_R(-\ell_2,4,3,-\ell_1)~.~~~\eean
Let us focus on $s_{12}$ cut, which appears only in
$A_{4,0}(1,2,3,4)$ and $A_{4,0}(1,3,4,2)$. The KK relation enables us
to fix two legs, so let us choose the basis with fixing 
$\ell_1,\ell_2$, then $A_L$ and $A_R$ in above expression are
already basis\footnote{We will not consider the BCJ relation in this
paper because the appearance of kinematic factors $s_{ij}$. We will
remark this point in the conclusion.}. It is clear that terms in
$A_{4,0}(1,2,3,4)$ and $A_{4,0}(1,3,4,2)$ are in different basis,
which of course can not be related by algebraic relations. Thus the
unitarity cut method tells us that there is no KK-like relation for
four-point one-loop primitive amplitudes.

%In four-point case these relations can be directly solved from
%$U(1)$-decoupling equations. But as we have analyzed, for $n\geq 6$
%the decoupling equations alone are not sufficient for solving
%$A_{n-m,m}$ when $m\geq 3$, while unitarity cut method can be
%applied to $n\geq 6$.

%%%%%%%%%%%%%%%%%%%%%%
\subsubsection{Proof of general case}
%%%%%%%%%%%%%%%%%%%%%

After the example of four-point, we move to general $n$-point
one-loop amplitude and try to prove (\ref{sublanswer}). The proof
will be given in following four steps. In the first step we will
identify all cuts of different color trace structures and prove the
color-ordering reversed identity (or reflection identity). In the
second step, we will discuss  example of  $A_{n-1,1}$ to warm up. In
the third step, we will present the proof of general case. In the
last step,  a technical detail will be explained.

%~\\

{\bf The first step:} We should identify all
cut contributions of a given trace structure.  For the partial
amplitudes $A_{n,0}$ (primitive amplitude), we have the following
equation:
\bea  A_{n,0}(1,2,\ldots,n)|_{cut}= \sum_{i=2}^{n-2}
\sum_{PCP\{1,2,\ldots,n\}}A_L(\ell_1,1,2,
\ldots,i,\ell_2)A_R(-\ell_2,i+1,\ldots,n,-\ell_1)+\{\ell_1\leftrightarrow\ell_2\}~,~~~\label{npan1}\eea
where $PCP\{\sigma\}$ is the {\sl partially cyclic permutation} of
$(1,2,...,n)$ such that particle $1$ is always at the
$A_L$\footnote{It is worth to notice that different $i$ will give
different $PCP\{\sigma\}$.}. For example, we have, for $i=3$,
\bea\sum_{PCP\{1,2,3,4,5\}}A_L(\ell_1,1,2,3,\ell_2)A_R(-\ell_2,4,5,-\ell_1)&=&A_L(\ell_1,1,2,3,\ell_2)A_R(-\ell_2,4,5,-\ell_1)\nonumber\\
&&+A_L(\ell_1,5,1,2,\ell_2)A_R(-\ell_2,3,4,-\ell_1)\nonumber\\
&&+A_L(\ell_1,4,5,1,\ell_2)A_R(-\ell_2,2,3,-\ell_1)~.~~~\nonumber\eea
Using the explicit expression (\ref{npan1}) we can show the
reflection identity mentioned in (\ref{Oneloop-reverse}). Using
reflection identity for tree-level amplitudes $A_L, A_R$ we get
\bea  A_{n,0}(1,2,\ldots,n)|_{cut}&=& \sum_{i=2}^{n-2}\sum_{PCP\{1,2,\ldots,n\}}(-)^{n+4}A_L(\ell_2,i,\ldots,2,1,\ell_1)A_R(-\ell_1,n,\ldots,i+1,-\ell_2)+\{\ell_1\leftrightarrow\ell_2\}\nonumber\\
&=&(-)^n\sum_{i=2}^{n-2}\sum_{PCP\{1,2,\ldots,n\}}
A_L(\ell_1,1,n,\ldots,n-i+2,\ell_2)A_R(-\ell_2,n-i+1,\ldots,2,-\ell_1)\nn
&&+\{\ell_1\leftrightarrow\ell_2\}~,~~~\nonumber\eea
where in the second line we have used the cyclic property under
$PCP$. Summing up all  terms in the second line  gives
$(-)^nA_{n,0}(1,n,\ldots,2)$, thus we have the reflection identity
for primitive amplitude as given in (\ref{Oneloop-reverse}).

For partial amplitudes with double trace structure, for example, the
$A_{c-1,n-c+1}(1,2,\ldots,c-1;c,\ldots,n)$, where $c\geq 2$, all double cuts are
\bea A_{c-1,n-c+1}(1,2,\ldots,c-1;c,\ldots,n)|_{cut}=\sum_{i\geq
k}^{n-c+k+1}\sum_{k=1}^{c-1}\sum_{PCP\{1,\ldots,c-1\}}
\sum_{CP\{c,\ldots,n\}}A_LA_R+
\{\ell_1\leftrightarrow\ell_2\}~,~~~\label{n-point-general-cut}\eea
where
\bea
A_LA_R=A_L(\ell_1,1,\ldots,k,\ell_2,n-i+k+1,\ldots,n)A_R(-\ell_2,k+1,
\ldots,c-1,-\ell_1,c,\ldots,n-i+k)~,~~~\nonumber\eea
and $CP\{\alpha\}$ is the {\sl cyclic permutation} over the set
$\a$. The difference between $PCP$ and $CP$ is that in $PCP$ we
require the particle $1$ is always at the $A_L$ to avoid the double
counting problem. Notice that $CP$ acts  on the set
$\{c,\ldots,n\}$, while $PCP$ acts on remaining set
$\{1,\ldots,c-1\}$.

Similarly to  $A_{n,0}$, there is also a reflection identity. By
accounting the $\{\ell_1\leftrightarrow\ell_2\}$ terms and using the
reflection identity for tree-level amplitudes $A_L$ and $A_R$, we get
\bea
A_LA_R+\{\ell_1\leftrightarrow\ell_2\}&=&(-)^{n+4}A_L(\ell_1,1,c-1,\ldots,c-k+1,\ell_2,n,\ldots,n-i+k+1)\nonumber\\
&&\times
A_R(-\ell_2,c-k,\ldots,2,-\ell_1,n-i+k,\ldots,c)+\{\ell_1\leftrightarrow\ell_2\}\nonumber\\
& = &
(-)^nA_{c-1,n-c+1}(1,c-1,\ldots,2;n-i+k,\ldots,c,n,\ldots,n-i+k+1)|_{cut}~.~~~\nonumber\eea
Using the cyclic permutation invariant of each trace, we  get the
reflection identity
\bea A_{c-1,n-c+1}(1,2,\ldots,c-1;c,\ldots,n)=(-)^n
A_{c-1,n-c+1}(c-1,c-2,\ldots,1;n,n-1,\ldots,c)~.~~~\label{Oneloop-double-ref}\eea

{\bf The second step:} As a warm
up, we consider the relation between $A_{n-1,1}$ and $A_{n,0}$. The
cuts of $A_{n-1,1}$ are
\bea A_{n-1,1}(1,2,\ldots,n-1;n)|_{cut}&=& \sum_{i=2}^{n-2}
\sum_{PCP\{1,2,\ldots,n-1\}}A_L(\ell_1,1,2,\ldots,i,\ell_2)
A_R(-\ell_2,i+1,\ldots,n-1,-\ell_1,n)\nonumber\\
&&+\sum_{i=2}^{n-2}\sum_{PCP\{1,2,\ldots,n-1\}}A_L(\ell_1,1,2,\ldots,i-1,\ell_2,n)A_R(-\ell_2,i,\ldots,n-1,-\ell_1)\nonumber\\
&&+\{\ell_1\leftrightarrow\ell_2\}~.~~~\label{npan2}\eea
Using KK relation to put $\ell_1, \ell_2$ at  two ends, we get
\bea &&A_{n-1,1}(1,2,\ldots,n-1;n)|_{cut}\nonumber\\&=&
-\sum_{i=2}^{n-2}\sum_{PCP\{1,2,\ldots,n-1\}}\sum_{OP\{i+1,\ldots,n-1\}\cup\{n\}}
A_L(\ell_1,1,2,\ldots,i,\ell_2)A_R(-\ell_2,i+1,\ldots,n,-\ell_1)\nonumber\\&&
-\sum_{i=2}^{n-2}\sum_{PCP\{1,2,\ldots,n-1\}}\sum_{
OP\{1,\ldots,i-1\}\cup\{n\}}A_L(\ell_1,1,\ldots,i-1,n,\ell_2)
A_R(-\ell_2,i,\ldots,n-1,-\ell_1)+\{\ell_1\leftrightarrow\ell_2\}\nonumber\\
&=&-\sum_{i=2}^{n-2}\sum_{PCP\{1,2,\ldots,n-1\}}\sum_{
OP\{1,\ldots,n-1\}\cup\{n\}}A_L(\ell_1,1,\ldots,i,\ell_2)A_R(-\ell_2,i+1,\ldots,n,-\ell_1)+\{\ell_1\leftrightarrow\ell_2\}~.~~~\nonumber\eea
%
%The result has a summation over partial cyclic permutation on
%$\{1,2,\ldots,n-1\}$ which inherits from $A_{n-1,1}$, while for
%$A_{n,0}$ the partial cyclic permutation operates on the whole label
%$\{1,2,\ldots,n\}$.
In order to express  above terms as primitive amplitudes, we
need to use following identity
\bea
&&\sum_{PCP\{1,2,\ldots,n-1\}}\sum_{OP\{1,\ldots,n-1\}\cup\{n\}}A_L(\ell_1,1,\ldots,i,\ell_2)A_R(-\ell_2,i+1,\ldots,n,-\ell_1)\nonumber\\
&&=\sum_{
OP\{2,3,\ldots,n-1\}\cup\{n\}}\sum_{PCP\{1,\ldots,n\}}A_L(\ell_1,1,\ldots,i,\ell_2)A_R(-\ell_2,i+1,\ldots,n,-\ell_1)~.~~~\label{id2to1}\eea
So the final result would be
\bea &&A_{n-1,1}(1,2,\ldots,n-1;n)|_{cut}\nn
&=& -\sum_{
OP\{2,3,\ldots,n-1\}\cup\{n\}}\sum_{i=2}^{n-2}\sum_{
PCP\{1,\ldots,n\}}A_L(\ell_1,1,\ldots,i,\ell_2)A_R(-\ell_2,i+1,\ldots,n,-\ell_1)\nonumber\\
&=&-\sum_{\alpha\in OP\{2,3,\ldots,n-1\}\cup\{n\}}A_{n,0}(1,\alpha)|_{cut}=
-\sum_{\beta\in~cyclic}A_{n,0}(\beta_1,\beta_2,\ldots,\beta_{n-1},n)|_{cut}~,~~~\eea
which is a special case of general formula (\ref{sublanswer}).

{\bf The third step:}  Now we consider the general case
(\ref{n-point-general-cut}). After using KK relation  to
$A_LA_R$, a typical term of (\ref{n-point-general-cut}) will become
\bea
& & (-1)^{n-c+1}\sum_{\sigma\in OP\{1,\ldots,k\}\cup\{n,\ldots,n-i+k+1\}}
A_L(\ell_1,\sigma(1,\ldots,k,n,...,n-i+k+1),\ell_2)\nonumber\\
&&\times \sum_{\W\sigma\in
OP\{k+1,\ldots,c-1\}\cup\{n-i+k,\ldots,c\}}A_R(-\ell_2,\W\sigma(k+1,\ldots,c-1,n-i+k,...,c),
-\ell_1)~,~~~\eea
where the ordering $\{c,c+1,\ldots,n\}$ has been reversed by the KK
relation. Other terms with given $k,i$ are obtained by cyclic
permutation of $k$-elements from set $\{1,2,\ldots,c-1\}$ and
$(i-k)$-elements from the set $\{n,\ldots,c\}$. Finally we need to sum
up all allowed  $k,i$. Regrouping them together, we can rewrite
$A_{c-1,n-c+1}|_{cut}$ as
\bea A_{c-1,n-c+1}(1,\ldots,c-1;c,\ldots,n)|_{cut}=(-1)^{n-c+1}
\sum_{i=2}^{n-2}\sum_{PCP\{1,\ldots,c-1\}}\sum_{CP\{c,\ldots,n\}}
\sum_{POP\{1,\ldots,c-1\}\cup\{n,\ldots,c\}}A_LA_R~,~\nn\eea
where
\bea
A_LA_R=A_L(\ell_1,1,\ldots,i,\ell_2)A_R(-\ell_2,i+1,\ldots,n,-\ell_1)~,~~~\nonumber\eea
and $POP\{\alpha\}\cup\{\beta\}$ (Partially Ordered Permutation) means ordered permutations between
sets $\{\alpha\}$ and $\{\beta\}$ while keeping $1$ in $A_L$. Using
identity
\bea
&&\sum_{PCP\{1,\ldots,c-1\}}\sum_{CP\{c,\ldots,n\}}\sum_{POP\{1,\ldots,c-1\}\cup\{n,\ldots,c\}}A_L(\ell_1,1,\ldots,i,\ell_2)A_R(-\ell_2,i+1,\ldots,n,-\ell_1)\nonumber\\
&=&\sum_{CP\{c,\ldots,n\}}\sum_{OP\{2,\ldots,c-1\}\cup\{n,\ldots,c\}}\sum_{PCP\{1,\ldots,n\}}A_L(\ell_1,1,\ldots,i,\ell_2)A_R(-\ell_2,i+1,\ldots,n,-\ell_1)~,~~~\label{idcto1}\eea
as well as (\ref{npan1}), $A_{c-1,n-c+1}|_{cut}$ can be simplified as
\bea
A_{c-1,n-c+1}(1,2,\ldots,c-1;c,\ldots,n)|_{cut}&=&(-1)^{n-c+1}
\sum_{CP\{c,\ldots,n\}}\sum_{OP\{2,\ldots,c-1\}\cup\{n,\ldots,c\}}A_{n,0}(1,2,\ldots,n)|_{cut}~.~~~\nn\eea
The two summations which are over all permutations between set
$\{1,2,\ldots,c-1\}$ and $\{n,\ldots,c\}$ with 1  fixed at the first
position, preserve the cyclic ordering of set $\{n,\ldots,c\}$.
They are  nothing but familiar
$$COP\{1,2,\ldots,c-1\}\cup\{n,\ldots,c\}~,$$ thus we finally prove
\bea A_{c-1,n-c+1}(1,2,\ldots,c-1;c,\ldots,n)&=&(-1)^{n-c+1}
\sum_{COP\{1,2,\ldots,c-1\}\cup\{n,\ldots,c\}}A_{n,0}(1,2,\ldots,n)~.~~~\eea
Using the  reflection identity (\ref{Oneloop-double-ref}) we can obtain
another form
\bea
A_{c-1,n-c+1}(1,2,\ldots,c-1;c,\ldots,n)&=&(-1)^{c-1}\sum_{
COP\{c-1,\ldots,1\}\cup\{c,c+1,\ldots,n\}}A_{n,0}(1,2,\ldots,n)~.~~~\eea

{\bf The last step:} The remaining thing we should clarify is the
identities (\ref{id2to1}) and (\ref{idcto1}). Since (\ref{id2to1})
is a special case of (\ref{idcto1}) when $c=n$, we just need to
prove the identity (\ref{idcto1}). To do so, we will consider terms
with leg 1 in the any given $k$-th position of  ordering in $A_LA_R$.
Since $k$ is chosen arbitrarily, if
the terms at both sides match up, then the identity is
true. Let us consider the summation of the first line in
(\ref{idcto1}),
\bea
\sum_{PCP\{1,\ldots,c-1\}}\sum_{CP\{c,\ldots,n\}}\sum_{POP\{1,\ldots,c-1\}\cup\{n,\ldots,c\}}A_L(\ell_1,1,\ldots,i,\ell_2)A_R(-\ell_2,i+1,\ldots,n,-\ell_1)~.~~~\eea
The  ordering of first summation and second summation does not
matter since they act on different sets. In order to hold leg $1$ in
the $k$-th position  in $A_L$, we should first take $POP$ action and
then $PCP$ action. The final result where  leg 1 is at the $k$-th
position is
\bea
\{1,2,\ldots,n\}\to\{OP\{\sigma_{k-1-m}\}\cup\{\sigma_m\},1,OP\{\sigma_{c+m-k-1}\}\cup\{\sigma_{n-c-m+1}\}\}~,~~~\label{idcto1p2}\eea
where
\bea
\{\sigma_{c+m-k-1},\sigma_{k-1-m}\}=\{2,\ldots,c-1\}~,~~\{\sigma_{m},\sigma_{n-c-m+1}\}=\{n,\ldots,c\}~.~~~\label{idcto1p2-t}\eea
The subscript of set $\sigma$ stands for the number of elements in
$\sigma$, and $m$ takes the value that all four $\sigma$ sets are
meaningful.

Then let us consider the summation of the second line in
(\ref{idcto1}),
\bea
\sum_{CP\{c,\ldots,n\}}\sum_{OP\{2,\ldots,c-1\}\cup\{n,\ldots,c\}}\sum_{PCP\{1,\ldots,n\}}A_L(\ell_1,1,\ldots,i,\ell_2)A_R(-\ell_2,i+1,\ldots,n,-\ell_1)~.~~~\eea
In order to hold leg 1 at $k$-th position, we should simply take the
following replacement using $PCP$,
\bea
\{1,2,\ldots,n\}\to\{n-k+2,\ldots,n,1,2,\ldots,n-k+1\}~.~~~\label{idcto1p1}\eea
Since actions under $POP$ and $CP$ will not change the position of
leg 1, we could then take the following replacements under $OP$
\bea
\{1,2,\ldots,n\}\to\{1,OP\{2,\ldots,c-1\}\cup\{n,\ldots,c\}\}~,~~~\eea
which means that $\{2,3,\ldots,n-k+1\}$ should be replaced by the
front $(n-k)$ elements of
$\{OP\{2,\ldots,c-1\}\cup\{n,\ldots,c\}\}$, and $\{n-k+2,\ldots,n\}$
should be replaced by the remaining $(k-1)$ elements of
$\{OP\{2,\ldots,c-1\}\cup\{n,\ldots,c\}\}$. By setting
\bea
\{\sigma'_{c+m-k-1},\sigma'_{k-1-m}\}=\{2,\ldots,c-1\}~,~~\{\sigma'_{n-c-m+1},\sigma'_{m}\}=\{n,\ldots,c\}~,~~~\label{idcto1p1-t}\eea
the above replacements can be compactly written as
\bea
&&\{2,\ldots,n-k+1\}\to\{OP\{\sigma'_{c+m-k-1}\}\cup\{\sigma'_{n-c-m+1}\}\}~,~~~\nonumber\\
&&\{n-k+2,\ldots,n\}\to\{OP\{\sigma'_{k-1-m}\}\cup\{\sigma'_{m}\}\}~.~~~\label{idcto1part2}\eea
The final result of actions (\ref{idcto1p1}) and (\ref{idcto1part2})
is
\bea \{1,2,\ldots,n\}\to\{OP\{\sigma'_{k-1-m}\}\cup\{\sigma'_{m}\},
1,OP\{\sigma'_{c+m-k-1}\}\cup\{\sigma'_{n-c-m+1}\}\}~.~~~\label{idcto1p3}\eea
Until now (\ref{idcto1p3}) is not equal to (\ref{idcto1p2}), since
we have $\{\sigma_{m},\sigma_{n-c-m+1}\}=\{n,\ldots,c\}$ in
(\ref{idcto1p2-t}) while
$\{\sigma'_{n-c-m+1},\sigma'_{m}\}=\{n,\ldots,c\}$ in
(\ref{idcto1p1-t}). Thus the elements in $\sigma_m$ and $\sigma'_m$
are different, and so are $\sigma_{n-c-m+1}$ and $\sigma'_{n-c-m+1}$.
But when considering sum of cyclic permutations
$\sum_{CP\{c,\ldots,n\}}$ at both sides, we can rewrite
$\{\sigma'_{n-c-m+1},\sigma'_{m}\}$ as
\bea
\{\sigma'_{n-c-m+1},\sigma'_{m}\}=\{n-m+2,\ldots,c+1,c;n,n-1,\ldots,n-m+1\}~,~~~
\eea
then we have $\sigma'_m=\sigma_m=\{n,n-1,\ldots,n-m+1\}$, thus
proving (\ref{idcto1}).

%%%%%%%%%%%%%%%%%%%%%%%%
\section{Partial amplitudes of two-loop amplitude}
%%%%%%%%%%%%%%%%%%%%%%%%%%

After one-loop calculation, we want to generalize our method to
higher loop. In this section, we will focus on two-loop case. The
color decomposition for two-loop amplitude in $U(N)$ gauge theory
%with all particles in the adjoint representation
can be
schematically written as
\bea \mathcal{A}^{2-loop}_n&=&\sum_{\sigma\in
S_n/Z_n}N_c^2\mbox{Tr}(\sigma_1,\ldots,\sigma_n)\left(A_{n}^{LC}(\sigma_1,\ldots,\sigma_n)+{1\over
N_c^2}A_{n}^{SC}(\sigma_1,\ldots,\sigma_n)\right)\nonumber\\
&&+\sum^{\lfloor{n/2}\rfloor}_{m=1}\sum_{\sigma\in
S_n/S_{n-m,m}}N_c\mbox{Tr}(\sigma_1,\ldots,\sigma_m)\mbox{Tr}(\sigma_{m+1},\ldots,\sigma_{n})A_{n-m,m}(\sigma_{1},\ldots,\sigma_m;\sigma_{m+1},\ldots,\sigma_{n})\nonumber\\
&&+\sum_{a=1}^{\lfloor{n/3}\rfloor}\sum_{(b-a)=a}^{\lfloor{(n-a)/2}\rfloor}\sum_{\sigma\in
S_n/S_{a,b-a,n-b}}\mbox{Tr}(\alpha)\mbox{Tr}(\beta)\mbox{Tr}(\gamma)A_{a,b-a,n-b}
(\alpha;\beta;\gamma)~,~~~\label{general-two-loop-color}\eea
where $\alpha=\{\sigma_1,\ldots,\sigma_a\}$,
$\beta=\{\sigma_{a+1},\ldots,\sigma_b\}$ and
$\gamma=\{\sigma_{b+1},\ldots,\sigma_n\}$. $S_{n-m,n}$ and
$S_{n-b,b-a,a}$ are corresponding groups that leaving the double
trace and triple trace invariant. The subscripts of partial
amplitudes denote the number of generators in traces. There are two
kinds of single trace structure: the ones with power $N_c^2$ are
leading-color single trace amplitudes and the others are
subleading-color single trace amplitudes, which come from
non-planar Feynman diagrams. The partial amplitudes are
 gauge invariant and may be calculated separately.

For two-loop case, there are  not many results on relations between
partial amplitudes,  due to the appearance of triple trace structure
as well as the subleading-color single trace structure, which make the
discussions more complicated. We would like to know, for example, if
 there are
relations  like (\ref{sublanswer}), so all other partial amplitudes
can be expressed by leading-color single trace partial amplitudes
$A_{n,0,0}$. If we can not achieve this goal, then how far  we can
go, i.e., what is the minimum basis of the partial amplitudes we
need  to completely determine the whole two-loop amplitudes. To
answer these questions, we would rely on both $U(1)$-decoupling
method and unitarity cut method.

The generalization of $U(1)$-decoupling method to two-loop is
straightforward, and the only difference from one-loop case is the
appearance of triple trace structure and subleading-color single trace
structure, which will lead to more decoupling equations. The solving
of all these equations is also  more complicated.

To generalize unitarity cut method to two-loop amplitude, we need to
introduce the triple cut. Then the whole two-loop amplitude becomes
(to prevent double-counting, we can fix leg $1$ in $A_L$)
\bea \mathcal{A}_n^{2-loop}|_{cut}=\sum_{states~of~\ell_1,\ell_2,\ell_3}\sum_{L,R}{\cal
A}_L^{full~tree}(\ell_1,\ell_2,\ell_3,\sigma_L) {\cal
A}_R^{full~tree}(-\ell_1,-\ell_2,-\ell_3,\sigma_R)~,~~~\label{two-loop-triple-cut}\eea
where  the summation is over all allowed triple cuts. Like one-loop case, there are also a few technical points we
need to point out. First for gauge theory,  we assume that  there is
no contribution by reduction process with only two inner propagators
( so  there is no triple cut available). It is the generalization of
the fact that there is no tadpole contribution at one-loop for gauge
theory. Secondly, we assume that there is basis for two-loop
amplitudes without the topology that two one-loop diagrams are
attached to each other at a vertex (such as the "bow-tie" diagram
given in \cite{Bern:2000dn}  or the "kissing box" diagrams given in
\cite{Bern:2008ap}). Since there is still no fully understanding of
basis of two loop amplitudes and how we can treat one basis to
another basis\footnote{There is a very nice paper
\cite{Gluza:2010ws} discussing the basis of planar two-loop
integrals.}, we can not show the assumption to be true, {\sl thus
our results in this section should be taken with caution up to this
uncertainty}. For the two loop MHV-amplitudes of ${\cal N}=4$
theory, Drummond and Henn \cite{Drummond:2010mb} have shown how to
tread the kissing box diagram to diagrams satisfying our assumption.
Thirdly we require there are at least two external gluons at $A_L$
and $A_R$, which is also reasonable for massless theory\footnote{One
argument for this is as follows. For one-loop case, the integration
$\int d^{D}\ell {1\over \ell^2 (\ell-k)^2}$ tell us that its form
should be $(k^2)^{D-4\over 2}$ by dimension analysis. Even with the
tensor structure in numerator, we will still see the appearing of
factor $k^2$ with proper power. This is one reason why massless
bubble gives zero contribution. For two loop with massless external
momenta such as ${f(\ell_i)\over \ell_1^2 \ell_2^2
(\ell_1+\ell_2+k)^2}$, similar consideration implies result
$(k^2)^{2D-6\over 2}$ for numerator $f(\ell)=1$, or $k_{\mu_1}
k_{\mu_2} ...k_{\mu_i} (k^2)^{{2D-6\over 2}+{n-i\over 2}}$ when
$f(\ell)$ is tensor structure. Thus dimensional regularization
implies the final result should be zero.}. Fourthly we assume our
triple cut discussion is true for general $(4-2\eps)$-dimension.
Otherwise our conclusion is true only for the ${\cal N}=4$ theory.
Our following discussions will base on above four technical
assumptions.

To see color structures of partial amplitudes coming from the triple
cut method, we do similar calculations as in
(\ref{one-loop-uni-trace})
\bea
&&\sum_{\ell_i}\left(\mbox{Tr}(\ell_1,\alpha_L,\ell_2,\beta_L,\ell_3,\gamma_L)+\mbox{Tr}(\ell_1,\W
\alpha_L,\ell_3,\W \beta_L,\ell_2,\W \gamma_L)\right)\nn
&&~~~~\times \left(
\mbox{Tr}(\ell_1,\alpha_R,\ell_2,\beta_R,\ell_3,\gamma_R)+\mbox{Tr}(\ell_1,\W
\alpha_R,\ell_3,\W \beta_R,\ell_2,\W
\gamma_R)\right)\nonumber\\
%&=&\mbox{Tr}(\ell_1,\alpha_L,\ell_2,\beta_L,\ell_3,\gamma_L)\mbox{Tr}(\ell_1,\alpha_R,\ell_2,\beta_R,\ell_3,\gamma_R)
%+\mbox{Tr}(\ell_1,\W \alpha_L,\ell_3,\W \beta_L,\ell_2,\W
%\gamma_L)\mbox{Tr}(\ell_1,\W \alpha_R,\ell_3,\W \beta_R,\ell_2,\W
%\gamma_R)\nonumber\\
%&&+\mbox{Tr}(\ell_1,\alpha_L,\ell_2,\beta_L,\ell_3,\gamma_L)\mbox{Tr}(\ell_1,\W
%\alpha_R,\ell_3,\W \beta_R,\ell_2,\W \gamma_R)+\mbox{Tr}(\ell_1,\W
%\alpha_L,\ell_3,\W \beta_L,\ell_2,\W
%\gamma_L)\mbox{Tr}(\ell_1,\alpha_R,\ell_2,\beta_R,\ell_3,\gamma_R)\nonumber\\
&=&\mbox{Tr}(  \gamma_L,\a_R,\b_L ,
 \gamma_R,\a_L ,\b_R )+\mbox{Tr}(  \W \b_L,\W \gamma_T, \W \a_L , \W \b_R,\W \gamma_L ,\W \a_R
)\nonumber\\
&&+\mbox{Tr}(\gamma_L,\W \a_R) Tr(\W \b_R,\b_L )Tr(\W \gamma_R,\a_L
)+\mbox{Tr}(\gamma_R ,\W \a_L)Tr( \W \b_L,\b_R) Tr(\W
\gamma_L,\a_R)~,~~~\label{2loop-triple-trace}\eea
which reproduces the familiar color structures given in
(\ref{general-two-loop-color}). The first two terms come from
$$\mbox{Tr}(\ell_1,\ldots,\ell_2,\ldots,\ell_3,\ldots)\mbox{Tr}(\ell_1,\ldots,\ell_2,\ldots,\ell_3,\ldots)~,~~~
\mbox{Tr}(\ell_1,\ldots,\ell_3,\ldots,\ell_2,\ldots)\mbox{Tr}(\ell_1,\ldots,\ell_3,\ldots,\ell_2,\ldots)~,~~~$$
which contribute to subleading-color single trace structure, while
the other two terms come from
$$\mbox{Tr}(\ell_1,\ldots,\ell_2,\ldots,\ell_3,\ldots)\mbox{Tr}(\ell_1,\ldots,\ell_3,\ldots,\ell_2,\ldots)~,~~~
\mbox{Tr}(\ell_1,\ldots,\ell_3,\ldots,\ell_2,\ldots)\mbox{Tr}(\ell_1,\ldots,\ell_2,\ldots,\ell_3,\ldots)~,~~~$$
which contribute to  leading-color single, double and triple trace
structures, depending on how many empty sets in these two terms are. It
is very important to notice that from  above discussions the pattern
of contributions to subleading-color single trace is different from
those  of other trace structures, so these two types will not mix
with each other in a simple way.

One simple result coming from the triple cut method is the
reflection identity for any type of partial amplitudes
\bea A_{a,b,n-a-b}(\a; \b; \gamma) = (-)^n A_{a,b,n-a-b}( \a^T;
\b^T;\gamma^T)~,~~~\label{2loop-ref} \eea
where $T$ means the reversing of ordering.

Having the experience of one-loop case, in this section, our
discussion will be more briefly. On the other hand, because the difficulty of the
problem, we have only some preliminary results and more works need
to be done in future.

%%%%%%%%%%%%%%%%%%%%
\subsection{Understanding four-point amplitude from $U(1)$-decoupling method}
%%%%%%%%%%%%%%%%%%%%%%%

Again we will start with the simplest example, i.e., the four-point
two loop amplitudes. We will use the $U(1)$-decoupling method in
this subsection and triple cut method in next subsection. It is
worth to remember that our discussion of $U(1)$-decoupling equation
is not new, and results in this subsection can be found,  for
example, in \cite{Bern:2002tk} (see also \cite{Bern:2010tq}). The
purpose of this subsection is to set up identities, so we can test
our generalized unitarity cut method in next subsection.

The color decomposition of four-point amplitude is
\cite{Bern:1997nh}
\bea \mathcal{A}_4^{2-loop}&=&\sum_{\sigma\in
S_4/Z_4}N_c^2\left(\mbox{Tr}(\sigma_1,\sigma_2,\sigma_3,\sigma_4)A_{4}^{LC}(\sigma_1,\sigma_2,\sigma_3,\sigma_4)+{1\over
N_c^2}\mbox{Tr}(\sigma_1,\sigma_2,\sigma_3,\sigma_4)A_{4}^{SC}(\sigma_1,\sigma_2,\sigma_3,\sigma_4)\right)\nonumber\\
&&+\sum_{\sigma \in
S_4/Z_3}N_c\mbox{Tr}(\sigma_1)\mbox{Tr}(\sigma_2,\sigma_3,\sigma_4)A_{1,3}(\sigma_1;\sigma_2,\sigma_3,\sigma_4)+\sum_{\sigma
\in
S_4/Z_2^3}N_c\mbox{Tr}(\sigma_1,\sigma_2)\mbox{Tr}(\sigma_3,\sigma_4)A_{2,2}(\sigma_1,\sigma_2;\sigma_3,\sigma_4)\nonumber\\
&&+\sum_{\sigma\in
S_4/Z^2_2}\mbox{Tr}(\sigma_1,\sigma_2)\mbox{Tr}(\sigma_3)\mbox{Tr}(\sigma_4)A_{2,1,1}(\sigma_1,\sigma_2;\sigma_3;\sigma_4)~,~~~\eea
where the summation for each color trace structure is over all
distinguished permutations, i.e., we should mod out permutations
making the color trace structure invariant.

There are  five kinds of trace structures: the subleading-color
single trace, the leading-color single trace, the double trace
$(3|1)$ and $(2|2)$, and finally the triple trace $(1|1|2)$. By
setting generators to be $U(1)$, subleading-color single trace can
never be mixed with  other color structures, so they have relations only
among themselves\footnote{Another way to see it is that we can take
$N_c$ as free parameter, so a function is zero when and only when
all coefficients of different $N_c$-power to be zero.}. For the
remaining color structures, by setting one generator to be $U(1)$,
they reduce to
\bea (4)\to (3)~;~~~(3|1)\to (3)~\mathrm{or}~(1|2)~;~~~(2|2)\to
(1|2)~;~~~(1|1|2)\to (1|2)~\mathrm{or}~(1|1|1)~.~~~\eea
Thus the reduced trace structure $(3)$ gives a relation between
$A_{4}^{LC}$ and $A_{1,3}$. The reduced  $(1|2)$ structure gives a
relation among $A_{1,3}, A_{2,2}$ and $A_{2,1,1}$ and finally the
reduced $(1|1|1)$ structure gives a relation of $A_{2,1,1}$.

More explicitly, by setting $T^4$ as $U(1)$, for the partial
amplitudes of subleading-color single trace, we get
\bea 0 & = & A_{4}^{SC} (4,1,2,3)+A_{4}^{SC} (4,3,1,2)+A_{4}^{SC}
(4,2,3,1) ~.~~~~\label{4-Sub-rel-1}\eea
It is worth to notice that it is exact the same form as tree-level
$U(1)$-decoupling equation. Then it is interesting to ask if there
is the same KK relation for subleading-color single partial
amplitudes? This question has no hint from $U(1)$-decoupling method,
but can be investigated by triple cut method in late subsection.

Let us continue to other $U(1)$-decoupling relation.   From the
reduced $N_c^2\mbox{Tr}(1,2,3)$ structure we can read out
\bea 0 & = & A_{1,3}(4;1,2,3)+\sum_{cyclic(123)}
A_{4}^{LC}(4,1,2,3)~,~~~\label{2loop-A4-1}\eea
so we can solve (other $A_{1,3}$  can be obtained simply by
relabeling)
\bea A_{1,3}(4;1,2,3)= -A_{4}^{LC} (4,1,2,3)- A_{4}^{LC}
(4,3,1,2)-A_{4}^{LC} (4,2,3,1) ~.~~~\label{2loop-A4-1trace}\eea
From the reduced $N_c \mbox{Tr}(1) \mbox{Tr}(2,3)$ structure we have
\bea 0 & = & A_{1,3}(1;
4,2,3)+A_{1,3}(1;4,3,2)+A_{2,2}(4,1;2,3)+A_{2,1,1}(2,3;1;4)~,~~\label{2loop-A4-2trace}
\eea
and finally from the reduced $\mbox{Tr}(1)\mbox{Tr}(2)\mbox{Tr}(3)$
structure, we have
\bea 0 & = &
A_{2,1,1}(4,1;2;3)+A_{1,2,1}(1;4,2;3)+A_{1,1,2}(1;2;4,3)~.~~\label{2loop-A4-3trace}\eea
Other independent relations will be obtained by relabeling of
indices.

Having these equations, we would like to ask if they are enough to
solve all the $A_{1,1,2}$ and $A_{2,2}$ in terms of $A_{4}^{LC}$.
Let us check this  by solving with (\ref{2loop-A4-3trace}) firstly.
There are $S_4/Z_2Z_2=6$ of $A_{1,1,2}$ and four equations, which can
be written as
\bean 0|_{T_4=1} & = & X_3+ X_5+
X_6,~~~~~0|_{T_2=1}=X_1+X_4+X_5,~~~0|_{T_3=1}=X_2+X_4+X_6,~~~0|_{T_1=1}=X_1+X_2+X_3~,~~~\eean
with
\bea X_1 & = & A_{2,1,1}(1,2;3;4)~,~~~X_2= A_{2,1,1}(1,3;2;4)~,~~~X_3=
A_{2,1,1}(1,4;2;3)~,~~~\nonumber\\
X_4 & = & A_{2,1,1}(2,3;1;4)~,~~~X_5= A_{2,1,1}(2,4;1;3)~,~~~X_6=
A_{2,1,1}(3,4;1;2)~.~~~\nonumber\eea
From these equations we can solve
\bea X_3=-X_1-X_2~,~~~X_4=-X_1-X_2~,~~~X_5=
X_2~,~~~X_6=X_1~,~~~\label{2loop-A4-3trace-1}\eea
where we have taken $X_1$ and $X_2$ as basis.  Putting them into
(\ref{2loop-A4-2trace}) we find solution for following three
 $A_{2,2}$:
\bea
Y_1=A_{2,2}(1,2;3,4)~,~~~Y_2=A_{2,2}(1,3;2,4)~,~~~Y_3=A_{2,2}(1,4;2,3)~~~~\eea
as
\bea &&Y_1=-X_1+\sum_{\sigma\in
COP\{1,2\}\cup\{3,4\}}A_{4}^{LC}(\sigma)~,~~~
Y_2=-X_2+\sum_{\sigma\in
COP\{1,3\}\cup\{2,4\}}A_{4}^{LC}(\sigma)~,~~~\nonumber\\
&&Y_3=X_1+X_2+\sum_{\sigma\in
COP\{1,4\}\cup\{2,3\}}A_{4}^{LC}(\sigma)~,~~~\label{2loop-A4-2trace-type2}\eea
where the difference between one loop $A_{2,2}$  and two loop
$A_{2,2}$ is the appearance of $A_{1,1,2}$ in
(\ref{2loop-A4-2trace-type2}).

%We can also substitute $A_{1,1,2}$ in (\ref{2loop-A4-3trace}) using
%(\ref{2loop-A4-2trace}), and get a relation among $A_{2,2}$ and
%$A_{4}^{LC}$ as
%
%\bea
%A_{2,2}(1,2;3,4)+A_{2,2}(1,3;2,4)+A_{2,2}(1,4;2,3)=3\sum_{\sigma\in
%S_4/Z_4}A_4^{LC}(\sigma)~.~~~\label{2loop-A4-2and1trace}\eea
%

In summary, from solving $U(1)$-decoupling equations, we see that
partial amplitudes of subleading-color trace structure are
themselves a special category, which has  same $U(1)$-decoupling
relation as the one for tree-level amplitudes. The remaining partial
amplitudes can be expressed as linear combination of all three
independent partial amplitudes of leading-color single trace
structure, plus two partial amplitudes of double (or triple) trace
structure.

%%%%%%%%%%%%%%%%%%%%
\subsection{Further understanding of four-point amplitude from unitarity cut method}
%%%%%%%%%%%%%%%%%%%%%%%

All  relations coming from $U(1)$-decoupling method in previous
subsection can be directly verified by unitarity cut method.
However, from one-loop example, we are warned that there are
non-trivial relations that can not be solved directly from
$U(1)$-decoupling relation. Thus we would like to ask are there any
more relations that are not revealed in $U(1)$-decoupling relation?
More specifically, we want to ask: (1) If we can express all partial
amplitudes of double or triple trace as a linear combination of partial amplitudes of
leading-color single trace? (2) If not, then we
would like to ask if the basis given in previous subsection, which
includes leading-color single trace and other two partial amplitudes
(it could  be two double (or triple) trace amplitudes), are
independent to each other.

In this subsection, we will discuss these problems using unitarity
cut method. Before going on, let us  work out the cut structures of
partial amplitudes. By a straightforward calculation, we can
express the cuts of the partial amplitude of leading-color single trace as
\bea
A_{4}^{LC}(1,2,3,4)|_{cut}&=&A_L(\ell_1,1,2,\ell_2,\ell_3)
A_R(-\ell_1,-\ell_3,-\ell_2,3,4)+A_L(\ell_1,4,1,\ell_2,\ell_3)A_R(-\ell_1,-\ell_3,-\ell_2,2,3)\nonumber\\
&&+ P\{\ell_1,\ell_2,\ell_3\}~,~~~\label{A4-lead}\eea
where $ P\{\ell_1,\ell_2,\ell_3\}$ means we should plus all the written terms with all the other permutations of
$\{\ell_1,\ell_2,\ell_3\}$. Similarly  for $A_{3,1}(1,2,3;4)$ we
have
\bea A_{3,1}(1,2,3;4)|_{cut}&=&A_L(\ell_1,1,2,\ell_2,\ell_3)A_R(-\ell_1,-\ell_3,4,-\ell_2,3)+A_L(\ell_1,1,2,\ell_2,\ell_3)A_R(-\ell_1,4,-\ell_3,-\ell_2,3)\nonumber\\
&&+A_L(\ell_1,3,1,\ell_2,\ell_3)A_R(-\ell_1,-\ell_3,4,-\ell_2,2)+A_L(\ell_1,3,1,\ell_2,\ell_3)A_R(-\ell_1,4,-\ell_3,-\ell_2,2)\nonumber\\
&&+A_L(\ell_1,1,\ell_2,4,\ell_3)A_R(-\ell_1,-\ell_3,-\ell_2,2,3)+A_L(\ell_1,1,\ell_2,\ell_3,4)A_R(-\ell_1,-\ell_3,-\ell_2,2,3)\nonumber\\
&&+ P\{\ell_1,\ell_2,\ell_3\}~,~~~\label{2l4pa412}\eea
and for $A_{2,2}(1,2;3,4)$,
\bea A_{2,2}(1,2;3,4)|_{cut}&=&A_L(\ell_1,\a_1,\a_2,\ell_2,\ell_3)A_R(-\ell_1,-\ell_3,\b_3,\b_4,-\ell_2)+A_L(\ell_1,\a_1,\a_2,\ell_2,\ell_3)A_R(-\ell_1,\b_3,\b_4,-\ell_3,-\ell_2)\nonumber\\
&&A_L(\ell_1,1,\ell_2,3,\ell_3)A_R(-\ell_1,-\ell_3,4,-\ell_2,2)+A_L(\ell_1,1,\ell_2,\ell_3,3)A_R(-\ell_1,4,-\ell_3,-\ell_2,2)\nonumber\\
&&A_L(\ell_1,1,\ell_2,4,\ell_3)A_R(-\ell_1,-\ell_3,3,-\ell_2,2)+A_L(\ell_1,1,\ell_2,\ell_3,4)A_R(-\ell_1,3,-\ell_3,-\ell_2,2)\nonumber\\
&&+ P\{\ell_1,\ell_2,\ell_3\}~,~~~\label{2l4pa413}\eea
where $\a,\b\in Z_2$. Finally the expression for the cuts of partial
amplitude of triple trace structure $A_{2,1,1}(1,2;3;4)$ is
\bea A_{2,1,1}(1,2;3;4)|_{cut}&=&A_L(\ell_1,\a_1,\a_2,\ell_2,\ell_3)A_R(-\ell_1,\b_3,-\ell_3,\b_4,-\ell_2)\nonumber\\
&&+A_L(\ell_1,1,\ell_2,\ell_3,3)A_R(-\ell_1,-\ell_3,4,-\ell_2,2)+A_L(\ell_1,1,\ell_2,3,\ell_3)A_R(-\ell_1,4,-\ell_3,-\ell_2,2)\nonumber\\
&&+A_L(\ell_1,1,\ell_2,\ell_3,4)A_R(-\ell_1,-\ell_3,3,-\ell_2,2)+A_L(\ell_1,1,\ell_2,4,\ell_3)A_R(-\ell_1,3,-\ell_3,-\ell_2,2)\nonumber\\
&&+ P\{\ell_1,\ell_2,\ell_3\}~,~~~\label{A4-12-3-4}\eea
where $\a, \b\in Z_2$. With these triple cut expressions, we can
check identities obtained from $U(1)$-decoupling equations. An
example is given in the Appendix B.

Having above settings, let us  study the first question by taking
$A_{2,1,1}(1,2;3;4)$ as an example. We want to express this
amplitude as\footnote{We have not assumed any relations between these
six amplitudes except the cyclic symmetry. There could be relations
and in fact, they do have ones as given in (\ref{2loop-ref}),  but it will not
affect the discussion here.}
\bean A_{2,1,1}(1,2;3;4) & = &x_1 A_4^{LC}(1,2,3,4)+ x_2
A_4^{LC}(1,2,4,3)+ x_3
A_4^{LC}(2,1,3,4)\nonumber\\
&&+ x_4 A_4^{LC}(2,1,4,3)+ x_5 A_4^{LC}(1,3,2,4)+ x_6
A_4^{LC}(1,4,2,3)~.~~~\nonumber\eean
Since $A_{2,1,1}(1,2;3;4)$ is symmetric under $1\leftrightarrow 2$
and $3\leftrightarrow 4$, we have $x_1=x_3$, $x_2=x_4$, $x_5=x_6$
and $x_1=x_2$, $x_3=x_4$, $x_5=x_6$. Furthermore, we know that
$A_{2,1,1}(1,2;3;4)=A_{1,1,2}(1;2;3,4)$, thus the exchanging of $(1,2)\leftrightarrow
(3,4)$ is also symmetric, this tells us $x_2=x_3$ and $x_5=x_6$.
Putting all these together we get
\bea A_{2,1,1}(1,2;3;4) & = & x \big(A_4^{LC}(1,2,3,4)+
A_4^{LC}(1,2,4,3)+
A_4^{LC}(2,1,3,4)+ A_4^{LC}(2,1,4,3)\big)\nonumber\\
&&+ y\big( A_4^{LC}(1,3,2,4)+
A_4^{LC}(1,4,2,3)\big)~.~~~\label{Test-1}\eea
Then the question becomes to find a solution $x,y$ for
(\ref{Test-1}). If identity (\ref{Test-1}) is true, it will be true
under unitarity cut. Writing down the cut expression as given in, for
example, (\ref{A4-12-3-4}) and (\ref{A4-lead}), at both sides and
comparing them, we could obtain equations for $x,y$. If there is
nonzero solution of $x,y$ to match up for all cuts, then there is a
relation, but if not, then $A_{2,1,1}$ can
not be expressed by $A_{4}^{LC}$.

Now we try to solve $x,y$ using the cut $s_{12}$. The contribution
for cut $s_{12}$ of $A_{2,1,1}(1,2;3;4)$ has been given in
(\ref{A4-12-3-4}). Let us use KK relation to take following six
amplitudes as basis for the left tree amplitudes:
\bea I_1 & = & A_L(\ell_1, 1,2,\ell_2, \ell_3)~,~~~ I_2=A_L(\ell_1,
2,1,\ell_2, \ell_3)~,~~~I_3=A_L(\ell_1, \ell_2,1,2, \ell_3)~,~~~\nn
I_4 & = & A_L(\ell_1, \ell_2,2,1, \ell_3)~,~~~I_5 =
A_L(\ell_1,1,\ell_2,2,\ell_3)~,~~~I_6 =
A_L(\ell_1,2,\ell_2,1,\ell_3)~,~~~\eea
and another six basis for the right tree amplitudes:
\bea K_1 & = & A_L(\ell_1, 3,4,\ell_2, \ell_3),~~ K_2=A_L(\ell_1,
4,3,\ell_2, \ell_3),~~K_3=A_L(\ell_1, \ell_2,3,4, \ell_3),\nn   K_4
& = & A_L(\ell_1, \ell_2,4,3, \ell_3),~~K_5  =
A_L(\ell_1,3,\ell_2,4,\ell_3),~~~K_6 =
A_L(\ell_1,4,\ell_2,3,\ell_3)~.~~~\eea
Then the coefficients of these $6\times 6$ basis for the left hand
side of (\ref{Test-1}) are given by
\bea &&(I_1+I_2)\times(K_3+K_4)\to-4~,~~~\nonumber\\
&&(I_3+I_4)\times(K_1+K_2+K_3+K_4+K_5+K_6)\to-4~,~~~\nonumber\\
&&(I_5+I_6)\times(K_1+K_2+K_3+K_4+K_5+K_6)\to-2~.~~~\nonumber\eea

For the right hand side of (\ref{Test-1}), amplitudes
$A_4^{LC}(1,2,3,4)$, $A_4^{LC}(1,2,4,3)$, $A_4^{LC}(2,1,3,4)$ and
$A_4^{LC}(2,1,4,1)$ will contribute to $s_{12}$ cut while
$A_4^{LC}(1,3,2,4)$ and $A_4^{LC}(1,4,2,3)$ will not. By expressing
$A_L$ and $A_R$ with basis $I_i, K_i$, we get coefficients of
these $6\times 6$ basis as
\bea &&(I_1+I_2)\times (K_1+K_2)\to 4x~,~~~\nonumber\\
&&(I_1+I_2)\times(K_3+K_4+K_5+K_6)\to 2x~,~~~\nonumber\\
&&(I_3+I_4)\times(K_1+K_2+K_5+K_6)\to 2x~,~~~\nonumber\\
&&(I_5+I_6)\times (K_1+K_2)\to 2x~.~~~\nonumber\eea
Comparing above two results it is obviously impossible to find
solution $x$, because even the basis at both sides do not match up!

Thus we have our first conclusion: {\sl we can not express the
double and triple trace partial amplitudes by leading single trace
partial amplitudes}. Although we have only done the four-point case,
we believe the conclusion is true for any $n$. Also we believe it is
 true for higher loops more than two.

There is one important point we want to remark. In our argument, we
have used the KK relation, but not the BCJ relation for tree-level
amplitudes, thus our conclusion is true only up to this
level. The reason we do not use BCJ relation is that in BCJ
relation, the kinematical factors involving $\ell_i$ will generally
appear, thus by unitarity cut method, coefficients of basis (we have
assumed there is one basis) will not be related to each other in simple
way and we will lose the predicability. We will come back to this
point in conclusion section.

Now we move to the second
question. From $U(1)$-decoupling method we know that without
considering partial amplitudes of subleading-color single trace, we
can express all the other partial amplitudes as a linear combination
of three independent $A_4^{LC}$ and two $A_{2,2}$ (or $A_{2,1,1}$).
To check if they are really independent, we wish to  find a
solution of $(\a,\b,x,y,z)$ so that
\bea \a A_{2,2}(1,3;2,4)+\b A_{2,2}(1,4;2,3)=x A_{4}^{LC}(1,2,3,4)+y
A_{4}^{LC}(1,2,4,3)+z A_{4}^{LC}(1,3,2,4)~.~~~\label{my-Test-2}\eea
Let us
focus on $s_{12}$ cut, and expand $A_L$ and $A_R$ in the above
basis $I_i, K_i$. The $A_{4}^{LC}(1,3,2,4)$'s in right hand side of
(\ref{my-Test-2}) do not contribute to $s_{12}$ cut. The
coefficients of $6\times 6$ basis for the left hand side of
(\ref{my-Test-2}) are
\bea &&(I_1+I_2+I_3+I_4)\times(K_1+K_2)\to
-4(\a+\b)~,~~~(I_1+I_2+I_3+I_4)\times(K_5+K_6)\to
-2(\a+\b)~,~~~\nonumber\\
&&(I_5+I_6)\times(K_1+K_2+K_3+K_4)\to -2(\a+\b)~,~~~I_5\times
K_5=I_6\times K_6=-6\a~,~~~\nonumber\\
&&I_5\times K_6=I_6\times K_5=-6\b~,~~~\nonumber\eea
while the coefficients for the right hand side of (\ref{my-Test-2})
are
\bea &&I_1\times K_1\to -2y~,~~~I_1\times K_2\to -2x~,~~~I_1\times
(K_3+K_5)\to -y~,~~~I_1\times (K_4+K_6)\to -x~,~~~\nonumber\\
&&I_2\times K_1\to -2x~,~~~I_2\times K_2\to -2y~,~~~I_2\times
(K_3+K_5)\to -x~,~~~I_2\times (K_4+K_6)\to -y~,~~~\nonumber\\
&&I_3\times K_4\to -2x~,~~~I_3\times K_3\to -2y~,~~~I_3\times
(K_2+K_6)\to -x~,~~~I_3\times (K_1+K_5)\to -y~,~~~\nonumber\\
&&I_4\times K_3\to -2x~,~~~I_4\times K_4\to -2y~,~~~I_4\times
(K_1+K_5)\to -x~,~~~I_4\times (K_2+K_6)\to -y~,~~~\nonumber\\
&&I_5\times(K_2+K_4+K_6)\to -x~,~~~I_5\times(K_1+K_3+K_5)\to
-y~,~~~\nonumber\\
&&I_6\times(K_2+K_4+K_6)\to -y~,~~~I_6\times(K_1+K_3+K_5)\to
-x~.~~~\nonumber\eea
All these basis should match up for a solution $(\a,\b,x,y,z)$.
However, noticing that  there are no $I_1\times K_3$ and $I_1\times
K_4$ terms in left hand side, it gives $x=y=0$, which leads further
to $\a=\b=0$. From this argument we see that there is no more
relation among three independent $A_4^{LC}$ and two $A_{2,2}$ (or
$A_{1,1,2}$). All these five partial amplitudes are indeed
independent to each other.

%%%%%%%%%%%%%%%%%%%%%
\subsection{KK-like relation for partial amplitudes of subleading-color single trace}
%%%%%%%%%%%%%%%%%%%%%
We have remarked in (\ref{4-Sub-rel-1}) that the $U(1)$-decoupling
relation for $A_{4}^{SC}$ is exactly the same as the one for tree-level amplitudes. For general $n$-point $A_n^{SC}$, we can also get
the same $U(1)$-decoupling relation using $U(1)$-decoupling method
\bea
\sum_{\sigma\in~cyclic}A_n^{SC}(\sigma_1,\sigma_2,\ldots,\sigma_{n-1},n)=0~,~~~\eea
where $T^n$ has been set to be $U(1)$. This similarity intrigues us
to ask if there is KK-like relation for $A_n^{SC}$. If the KK
relation is true for $A_n^{SC}$, the independent partial amplitudes
of subleading-color trace will be greatly reduced from $(n-1)!$ to
$(n-2)!$. Since KK relation can not be derived from
$U(1)$-decoupling method, we need to  investigate this problem by
unitarity cut method.

It is worth to mention that the reflection identity and
$U(1)$-decoupling identity, which have been shown to be true, are
special cases of KK relation.  The first non-trivial KK relation,
i.e., KK relation that is different from $U(1)$-decoupling and
reflection relation, appears in six-point case. For example, we can
write down
%For KK relation of $A_6^{SC}$, we can take the basis
%that leg 1 and 6 are  fixed at the first and last positions
%respectively, and express all  other amplitudes as linear
%combination of the basis. Let us consider following non-trivial KK
%relation
%
\bea
A_6^{SC}(1,2,3,6,4,5)&=&A^{SC}_6(1,2,3,5,4,6)+A^{SC}_6(1,2,5,3,4,6)+A^{SC}_6(1,2,5,4,3,6)\nonumber\\
&&+A^{SC}_6(1,5,2,3,4,6)+A^{SC}_6(1,5,2,4,3,6)+A^{SC}_6(1,5,4,2,3,6)~.~~~\label{6point-KK}\eea
If above  relation is true, it should be true for every triple cut.
At  first sight it seems obscure, since all  terms under the triple
cut will have the patterns
\bea
A_L(\ell_1,\ldots,\ell_2,\ldots,\ell_3)A_R(-\ell_1,\ldots,-\ell_2,\ldots,-\ell_3)~,~~~
A_L(\ell_1,\ldots,\ell_3,\ldots,\ell_2)A_R(-\ell_1,\ldots,-\ell_3,\ldots,-\ell_2)~,~~~\eea
which are hard to observe relations among them. The matching  of
every cut in left and right hand sides of (\ref{6point-KK}) is quite
non-trivial.

There are totally twenty-five different cuts\footnote{For general
$n$, there are $2^{n-1}-(n+1)$ different cuts to be considered.}
$s_{1i}$, $s_{1ij}$ and $s_{1ijk}$ for (\ref{6point-KK}): fifteen
two-particle cuts and ten three-particle cuts. Equation
(\ref{6point-KK}) has symmetries $\{2\leftrightarrow 5,
3\leftrightarrow 4\}$ and $ \{1\leftrightarrow 6, 4\leftrightarrow
2, 5\leftrightarrow 3\}$, thus many cuts can be related to each
other and we need to check only one cut for each orbit given by
symmetry group. With this consideration, cuts to be checked are
reduced to the following eleven: six two-particle cuts
\bea s_{12}\sim s_{15}\sim s_{46}\sim s_{36}~;~~~s_{13}\sim
s_{14}\sim s_{56}\sim s_{26}~;~~~s_{45}\sim s_{23}~;~~~s_{34}\sim
s_{25}~;~~~s_{35}\sim s_{24}~;~~~s_{16}~,~~~\nonumber\eea
and five three-particle cuts
\bea s_{126}\sim s_{156}\sim s_{136}\sim s_{146}~;~~~s_{123}\sim
s_{145}~;~~~s_{124}\sim
s_{135}~;~~~s_{125}~;~~~s_{134}~.~~~\nonumber\eea

As an example of how terms match up, we consider cut $s_{134}$,
where the expression for  $A_6^{SC}|_{cut}$   is  relatively
simple. For the left hand side of (\ref{6point-KK}), we have
\bea A^{SC}_6(1,2,3,6,4,5)|_{cut}=
A_L(\ell_1,1,\ell_2,4,\ell_3,3)A_R(-\ell_1,6,-\ell_2,2,-\ell_3,5)+
P\{\ell_1,\ell_2,\ell_3\}~.~~~\nonumber\eea
For the right hand side,  six $A_6^{SC}|_{cut}$'s have  following
contributions:
\bea &&A^{SC}_6(1,2,3,5,4,6)|_{cut}=
A_L(\ell_1,1,\ell_2,4,\ell_3,3)A_R(-\ell_1,5,-\ell_2,2,-\ell_3,6)+
P\{\ell_1,\ell_2,\ell_3\}\nn
 &&A^{SC}_6(1,2,5,3,4,6)|_{cut}=\nonumber\\
&&A_L(\ell_1,1,\ell_2,3,4,\ell_3)\big(A_R(-\ell_1,2,5,-\ell_2,-\ell_3,6)+A_R(-\ell_1,-\ell_2,2,5,-\ell_3,6)+A_R(-\ell_1,5,-\ell_2,2,-\ell_3,6)\big)\nonumber\\
&&+A_L(\ell_1,1,\ell_2,\ell_3,3,4)\big(A_R(-\ell_1,6,-\ell_2,2,5,-\ell_3)+A_R(-\ell_1,-\ell_2,2,5,-\ell_3,6)\big)\nonumber\\
&&+A_L(\ell_1,1,\ell_2,4,\ell_3,3)A_R(-\ell_1,-\ell_2,2,5,-\ell_3,6)+
P\{\ell_1,\ell_2,\ell_3\}\big)~,~~~\nonumber\eea
and remaining  four partial amplitudes have
\bea A^{SC}_6(1,2,5,4,3,6)|_{cut}&=&
(A^{SC}_6(1,2,5,3,4,6)|_{cut})|_{3\leftrightarrow4};\nn
A^{SC}_6(1,5,2,4,3,6)|_{cut}&=&
(A^{SC}_6(1,2,5,3,4,6)|_{cut})|_{(3\leftrightarrow4,2\leftrightarrow5)};\nonumber\\
A^{SC}_6(1,5,2,3,4,6)|_{cut}&=&
(A^{SC}_6(1,2,5,3,4,6)|_{cut})|_{2\leftrightarrow5};\nn
A^{SC}_6(1,5,4,2,3,6)|_{cut}&=&
(A^{SC}_6(1,2,3,5,4,6)|_{cut})|_{(3\leftrightarrow4,2\leftrightarrow5)}.\nonumber\eea
To compare terms at both sides, we expand them into a chosen basis,
i.e., the basis independent to each other up to KK relation.
%In
%order to compare terms in left hand side with terms in right hand
%side, we should expand tree amplitudes into basis. The basis are
%given by fixing two legs using KK relation.
The choice we have made here is that  leg 3 and 4 of $A_L$-part are
at the first and last position respectively, while leg 2 and  leg 5 of
$A_R$-part are at  the first and last position respectively. Thus we need to
compare coefficients of  $24\times 24$ basis at both sides.

 In order to make an
impression of how these basis match up, we give some details. When we
expand six $A^{SC}$'s at the right hand side of (\ref{6point-KK}) by
the basis, we will get following expression 
%plus its all permutations $P\{\ell_1,\ell_2,\ell_3\}$:
%
\bea  & & RHS|_{cut}= A_L(3,\ell_1,1,\ell_2,\ell_3,4)\times
\Big(2A_R(2,-\ell_3,6,-\ell_2,-\ell_1,5)+2A_R(2,-\ell_3,6,-\ell_1,-\ell_2,5)\nn
&&+A_R(2,-\ell_2,6,-\ell_1,-\ell_3,5)-A_R(2,-\ell_2,6,-\ell_3,-\ell_1,5)+A_R(2,-\ell_2,-\ell_3,6,-\ell_1,5)+A_R(2,-\ell_3,-\ell_2,6,-\ell_1,5)\nn
&&+A_R(2,-\ell_2,-\ell_1,6,-\ell_3,5)+A_R(2,-\ell_3,-\ell_1,6,-\ell_2,5)-A_R(2,-\ell_1,-\ell_2,6,-\ell_3,5)+A_R(2,-\ell_1,-\ell_3,6,-\ell_2,5)\Big)\nonumber\\
&&+A_L(3,\ell_1,1,\ell_3,\ell_2,4)\times\Big(2A_R(2,-\ell_3,6,-\ell_2,-\ell_1,5)-A_R(2,-\ell_2,6,-\ell_1,-\ell_3,5)-A_R(2,-\ell_2,6,-\ell_3,-\ell_1,5)\nonumber\\
&&+A_R(2,-\ell_2,-\ell_3,6,-\ell_1,5)+A_R(2,-\ell_3,-\ell_2,6,-\ell_1,5)-A_R(2,-\ell_1,-\ell_2,6,-\ell_3,5)+A_R(2,-\ell_1,-\ell_3,6,-\ell_2,5)\nonumber\\
&&-A_R(2,-\ell_2,-\ell_1,6,-\ell_3,5)-A_R(2,-\ell_3,-\ell_1,6,-\ell_2,5)\Big)+A_L(3,\ell_3,\ell_1,1,\ell_2,4)\times\Big(-2A_R(2,-\ell_1,6,-\ell_2,-\ell_3,5)\nonumber\\
&&+A_R(2,-\ell_2,6,-\ell_3,-\ell_1,5)+A_R(2,-\ell_2,6,-\ell_1,-\ell_3,5)+A_R(2,-\ell_3,-\ell_2,6,-\ell_1,5)+A_R(2,-\ell_2,-\ell_3,6,-\ell_1,5)\nonumber\\
&&-2A_R(2,-\ell_1,-\ell_2,6,-\ell_3,5)\Big)+A_L(3,\ell_1,\ell_3,1,\ell_2,4)\times\Big(2A_R(2,-\ell_3,6,-\ell_2,-\ell_1,5)+A_R(2,-\ell_2,6,-\ell_3,-\ell_1,5)\nonumber\\
&&+A_R(2,-\ell_2,6,-\ell_1,-\ell_3,5)+3A_R(2,-\ell_3,-\ell_2,6,-\ell_1,5)
+A_R(2,-\ell_2,-\ell_3,6,-\ell_1,5)\Big)+P\{\ell_1,\ell_2,\ell_3\}~.~~~\label{R-part}\eea
%
%This result is difficult to deal with, and no clear cancelation can
%be seen within these terms. This is different from the corresponding
%terms in left hand side, which has only several simple terms
Similarly, the expansion of left hand side of (\ref{6point-KK})  is
equal to following expression
% plus its all permutations
%$P\{\ell_1,\ell_2,\ell_3\}$:
%
\bea
 LHS|_{cut}&=&\Big(A_L(3,\ell_1,1,\ell_2,\ell_3,4)+A_L(3,\ell_1,1,\ell_3,\ell_2,4)
 +A_L(3,\ell_3,\ell_1,1,\ell_2,4)+A_L(3,\ell_1,\ell_3,1,\ell_2,4)\Big)\nn
&&\times
\Big(A_R(2,-\ell_2,6,-\ell_3,-\ell_1,5)+A_R(2,-\ell_2,6,-\ell_1,-\ell_3,5)+A_R(2,-\ell_2,-\ell_3,6,-\ell_1,5)\nn
&&+A_R(2,-\ell_3,-\ell_2,6,-\ell_1,5)\Big)+P\{\ell_1,\ell_2,\ell_3\}~.\label{L-part}\eea
From above expressions, it is very difficult to see that $RHS|_{cut}$ will
equal to $LHS|_{cut}$. But if we explicitly write down the terms in $P\{\ell_1,\ell_2,\ell_3\}$ of $RHS|_{cut}$, we will see much cancellations. For example, if we only  explicitly write down the terms with exchange of $\ell_2$ and $\ell_3$ in $P\{\ell_1,\ell_2,\ell_3\}$ of $RHS|_{cut}$, the expression will be highly reduced to 
%However, it is amazing that when $R$ plus terms with permutation
%$\ell_2\leftrightarrow\ell_3$, many cancelations happen and we get a
%simple result as
%
\bea  && RHS|_{cut}=\nn
&&\Big(A_L(3,\ell_1,1,\ell_2,\ell_3,4)+A_L(3,\ell_1,1,\ell_3,\ell_2,4)\Big)\times
\Big(2A_R(2,-\ell_2,-\ell_3,6,-\ell_1,5)+2A_R(2,-\ell_3,-\ell_2,6,-\ell_1,5)\nonumber\\
&&+A_R(2,-\ell_3,6,-\ell_2,-\ell_1,5)+A_R(2,-\ell_3,6,-\ell_1,-\ell_2,5)+A_R(2,-\ell_2,6,-\ell_3,-\ell_1,5)+A_R(2,-\ell_2,6,-\ell_1,-\ell_3,5)\Big)\nonumber\\
&&+\Big(A_L(3,\ell_3,\ell_1,1,\ell_2,4)+A_L(3,\ell_1,\ell_3,1,\ell_2,4)\Big)\times
\Big(2A_R(2,-\ell_2,6,-\ell_3,-\ell_1,5)+2A_R(2,-\ell_2,6,-\ell_1,-\ell_3,5)\nonumber\\
&&+A_R(2,-\ell_2,-\ell_3,6,-\ell_1,5)+A_R(2,-\ell_3,-\ell_2,6,-\ell_1,5)
+A_R(2,-\ell_2,-\ell_1,6,-\ell_3,5)+A_R(2,-\ell_1,-\ell_2,6,-\ell_3,5)\Big)\nn
&&+CP\{\ell_1,\ell_2,\ell_3\}~,~~\eea
which can be easily seen equal to $LHS|_{cut}$.
%matches up to $(L+ L|_{\ell_2\leftrightarrow\ell_3})$. In
%other words, six permutations have been divided into three groups
%and for each group, the left and right hand sides  will match up
%after above nontrivial cancelations.

%The check of cut $s_{134}$ is relatively simpler as explained above.
Unlike the simplicity of the cut $s_{134}$, other cuts are more difficult to check. We have implemented it in
{\bf Mathematica} and found that  for all cuts $s_{1i}$, $s_{1ij}$
and $s_{1ijk}$, after using the KK relation of tree-level
amplitudes, the equation (\ref{6point-KK}) always holds.

Beyond six-point, we have also checked the case of seven-point
 by {\bf Mathematica} and the complexity increases dramatically
with the increasing of $n$. For seven points, the KK relation is
also true.

Next is the eight-point case, but we found that, by checking several cuts,
\bea A^{SC}(1, \{ 2, 3\}, 8,\{4,5,6,7\}) \neq \sum_{\sigma\in OP
\{2,3\}\bigcup\{ 7,6,5,4\}} A^{SC}(1,\sigma, 8)~,~~~\eea
where "$\neq$" means under the cut, the left hand side is not equal to the right
hand side. However, we do find that
\bea & & A^{SC}(1, \{ 2, 3\}, 8,\{4,5,6,7\})+ A^{SC}(1, \{ 3, 2\},
8,\{4,5,6,7\})\nn & = & \sum_{\sigma\in OP \{2,3\}\bigcup\{
7,6,5,4\}} A^{SC}(1,\sigma, 8)+\sum_{\sigma\in OP \{3,2\}\bigcup\{
7,6,5,4\}} A^{SC}(1,\sigma, 8)~,~~~\label{8-KK-2-4} \eea
where "$=$" means for all cuts the both sides are equal. For
another KK relation of $A^{SC}(1, \{ 2, 3,4\}, 8,\{5,6,7\})$, it is
also not true by unitarity cut method, but we find that
\bea \sum_{cyclic \{2,3,4\}}A^{SC}(1, \{ 2, 3,4\}, 8,\{5,6,7\}) =
-\sum_{cyclic \{2,3,4\}}\sum_{\sigma\in OP \{2,3,4\}\bigcup\{
7,6,5\}} A^{SC}(1,\sigma, 8)~~~~\label{8-KK-3-3}\eea
is true under all triple cuts. It is also strange to find that the
relation
\bea \sum_{cyclic \{2,3,4\}}A^{SC}(1, \{ 2, 3,4,5\}, 8,\{6,7\}) =
\sum_{cyclic \{2,3,4\}}\sum_{\sigma\in OP \{2,3,4,5\}\bigcup\{
7,6\}} A^{SC}(1,\sigma, 8)~~~~\eea
is true under all triple cuts. 
%This is different from
%(\ref{8-KK-2-4}) and (\ref{8-KK-3-3}), where we have added the
%partial amplitudes with cyclic permutations on set $\a$ so that they
%would be true seen from unitarity cut method.
 The case of nine point
is too complicated even for the computer.

The observation of eight point is very mysterious for us and we do
not  understand why naive KK relation fails for higher points. It is
possible that KK relation is true for higher points, but our triple
cut method can not assure it. In other words, although the
integrands at both sides do not match up under our unitarity cut
method, the final integrated results may match up. We are continuing
the investigation of this problem.

%%%%%%%%%%%%%%%%%%%%%
\section{Conclusion}
%%%%%%%%%%%%%%%%%%%%%%

In this paper we have used the unitarity cut method
\cite{Landau:1959fi,Bern:1994cg} to study relations among
color-ordered partial amplitudes of gauge theory at one-loop and
two-loop. At one-loop we have proved the known result
(\ref{sublanswer}) that partial amplitudes of double trace structure
can be completely solved as a linear combination of primitive
amplitudes \cite{Bern:1994zx} by using KK relation of tree-level
amplitudes. Our proof gives a clear physical picture for the
similarity between relation (\ref{sublanswer}) and tree-level KK
relation  (\ref{KK}). The reflection identity of any-loop amplitudes
can also be understood explicitly from reflection identity of tree
amplitudes by unitarity cut method although it can also be
understood directly from the pure group property of gauge theory.

At two-loop level, unitarity cut method has also helped us to
understand several interesting questions. First it is shown that
just partial amplitudes of leading-color single trace structure are
not enough to solve partial amplitudes of other trace structures.
This can also be understood by noticing that leading-color partial
amplitudes include only planar diagrams\footnote{We would like to
thank referee for several enlightening remarks.}. Then the unitarity
cut method leads us to the possibility that there is KK-like
relation for partial amplitudes of subleading-color single trace
structure, where examples of six-, seven- and eight-point, have
been explicitly studied.

Our result in this paper is only one little step of the application
of unitarity cut method for understanding  the relation of loop amplitudes.
There are many things which are still not clear and should be discussed in
future.

The first thing we want to understand more is the role of tree-level
BCJ relation for loop amplitudes. In this paper, we have used only
tree-level KK relation and have deliberately avoided the use of BCJ
relation. The main reason is that BCJ relation will involve the
kinematic factors $s_{\ell_i i}$, which makes the discussion in the
frame of unitarity cut method very complicated. The generalization
of BCJ relation to loop-level has been discussed in
\cite{BjerrumBohr:2011xe,Bern:2010ue,Bern:2010yg,Bern:2011ia}, where
not the whole partial amplitude, but some parts of it have relations. The correspondence of this point in the unitarity cut
method is following: we may get match up for some cuts, but not for
all cuts. Thus we do not get the relation for the whole amplitude, but
do get relations for  parts of amplitude detected by these matching
cuts. Of course, many works are needed to make above picture clear.

The second thing worth to do is to systematically study two-loop
partial amplitudes. The mysterious KK-like  relation for subleading-color
single trace partial amplitudes has not been understood. The
similarity has also intrigued us to ask the possibility of BCJ-like
relation for subleading-color single  trace partial amplitudes.
 Moreover,
although the basis found by $U(1)$-decoupling method in four-point
case is the same basis found by unitarity cut method, we are not
sure if this will be true for general $n$. Just like  one-loop
example, (\ref{sublanswer}) reduces to $U(1)$-decoupling equation
for $n\leq 5$, but is different for $n\geq 6$.

It is also interesting to use unitarity cut method to discuss
partial amplitudes for more than two loops. With the increasing of
loops, the complexity will also increase a lot, thus a better idea to
implement this method would be welcome.

%%%%%%%%%%%%%%%%%%%%%%
\subsection*{Acknowledgements}
%%%%%%%%%%%%%%%%%%%%%%

We  would like to thank Rutger Boels for collaboration at early
stage of this project and the hospitality of KITPC, China, where
final part of this work was done. We are supported by fund from
Qiu-Shi, the Fundamental Research Funds for the Central Universities
with contract number 2010QNA3015, as well as Chinese NSF funding
under contract No.10875104, No.11031005.
%%%%%%%%%%%%%%%%%%%%
\appendix
%%%%%%%%%%%%%%%%%%%%%

%%%%%%%%%%%%%%%%%
\section{Direct verification of relations for two-loop four-point amplitude}
%%%%%%%%%%%%%%%%%%%%

Two-loop four-gluon partial amplitudes of $SU(N)$ $\mathcal{N}=4$
super-Yang-Mills theory have been computed in \cite{Bern:1997nh} by
cut method. We would like to verify the relations of two-loop four-point
amplitudes directly using these results.

The relations we have obtained are
\bea 0 & = & A_{4}^{SC} (1,2,3,4)+A_{4}^{SC} (1,2,4,3)+A_{4}^{SC}
(1,4,2,3)~,~~~\label{appendix-test-1}\eea
and
\bea
A_{2,2}(1,2;3,4)+A_{2,2}(1,3;2,4)+A_{2,2}(1,4;2,3)=3\sum_{\sigma\in
S_4/Z_4}A_4^{LC}(\sigma)~.~~~\label{appendix-test-2}\eea
In order to verify these two relations, we need to know the
corresponding partial amplitudes. In \cite{Bern:1997nh} partial
amplitudes are given as a linear combination of some planar and
non-planar basis. Written in our convention, the
leading-color single trace amplitude is
\bea A_4^{LC}(1,2,3,4)=A_4^P(1,2;3,4)+A_4^P(1,4;3,2)~,~~~\eea
and the subleading-color single trace amplitude is
\bea
A_4^{SC}(1,2,3,4)&=&2A_4^P(1,2;3,4)+2A_4^P(1,2;4,3)+2A_4^P(1,4;2,3)+2A_4^P(1,4;3,2)\nonumber\\
&&-4A_4^P(1,3;2,4)-4A_4^P(1,3;4,2)+2A_4^{NP}(1,2;3,4)+2A_4^{NP}(1,2;4,3)\nonumber\\
&&+2A_4^{NP}(1,4;2,3)+2A_4^{NP}(1,4;3,2)-4A_4^{NP}(1,3;2,4)-4A_4^{NP}(1,3;4,2)~,~~~\eea
then the double trace amplitude is
\bea
A_{2,2}(1,2;3,4)&=&6A_4^{P}(1,2;3,4)+6A_4^{P}(1,2;4,3)+4A_4^{NP}(1,2;3,4)+4A_4^{NP}(1,2;4,3)\nonumber\\
&&-2A_4^{NP}(1,4;2,3)-2A_4^{NP}(1,4;3,2)-2A_4^{NP}(1,3;2,4)-2A_4^{NP}(1,3;4,2)~,~~~\eea
where $A^P$ and $A^{NP}$ are functions of two-loop planar and non-planar
scalar double-box integrals as defined in \cite{Bern:1997nh}.
%
%
%We have using the fact that
%
%\bea
%A_4^{P~or~NP}(1,2;3,4)&\equiv&-s_{12}^2s_{23}A_4^{tree}(1,2,3,4)I_4^{P~or~NP}(s_{12},s_{23})\nonumber\\
%&=&-s_{34}^2s_{14}A_4^{tree}(3,4,1,2)I_4^{P~or~NP}(s_{34},s_{41})\equiv
%A_4^{P~or~NP}(3,4;1,2)\nonumber\\
%&=&-s_{21}^2s_{14}A_4^{tree}(2,1,4,3)I_4^{P~or~NP}(s_{21},s_{14})\equiv
%A_4^{P~or~NP}(2,1;4,3)\nonumber~~~\eea
%
%to rewrite all $A_4^{P}$ and $A_4^{NP}$ in six basis respectively.

Firstly let us verify (\ref{appendix-test-1}). The coefficients of
each basis can be directly written down as
\bea \begin{array}{c|c|c|c|c|c|c}
   & A_4^P(1,2;3,4) & A_4^P(1,2;4,3) & A_4^P(1,3;2,4) & A_4^P(1,3;4,2) & A_4^P(1,4;2,3)& A_4^P(1,4;3,2) \\ \hline
  A_{4}^{SC}(1,2,3,4)& 2 & 2 & -4 & -4 & 2 & 2 \\ \hline
  A_{4}^{SC}(1,2,4,3)& 2 & 2 & 2 & 2 & -4 & -4 \\ \hline
  A_{4}^{SC}(1,4,2,3)& -4 & -4 & 2 & 2 & 2 & 2
  \end{array}~~~.\nonumber\eea
It is clear to see that sum of each basis is zero. The sum for
non-planar basis is the same as planar basis, thus verifing
(\ref{appendix-test-1}).

Then we continue to verify (\ref{appendix-test-2}). We
consider planar basis, and for the left hand side we have
\bea
\begin{array}{c|c|c|c|c|c|c}
   & A_4^P(1,2;3,4) & A_4^P(1,2;4,3) & A_4^P(1,3;2,4) & A_4^P(1,3;4,2) & A_4^P(1,4;2,3)& A_4^P(1,4;3,2) \\ \hline
  A_{2,2}(1,2;3,4)& 6 & 6 & 0 & 0 & 0 & 0 \\  \hline
  A_{2,2}(1,3;2,4)& 0 & 0 & 6 & 6 & 0 & 0 \\ \hline
  A_{2,2}(1,4;2,3)& 0 & 0 & 0 & 0 & 6 & 6
\end{array}~~~.\nonumber\eea
The coefficient for each two-loop planar basis is six. It is easy to
get the coefficient for each basis of the right hand side, which is also
six. 

%and for the right hand side, we have
%
%\bea \begin{array}{ccccccc}
%   & A_4^P(1,2;3,4) & A_4^P(1,2;4,3) & A_4^P(1,3;2,4) & A_4^P(1,3;4,2) & A_4^P(1,4;2,3)& A_4^P(1,4;3,2) \\
%  3A_{4}^{LC}(1,2,3,4)& 3 & 0 & 0 & 0 & 0 & 3 \\
%  3A_{4}^{LC}(1,2,4,3)& 0 & 3 & 0 & 3 & 0 & 0 \\
%  3A_{4}^{LC}(1,3,2,4)& 0 & 0 & 3 & 0 & 3 & 0 \\
%  3A_{4}^{LC}(1,4,2,3)& 0 & 0 & 3 & 0 & 3 & 0 \\
%  3A_{4}^{LC}(1,3,4,2)& 0 & 3 & 0 & 3 & 0 & 0 \\
%  3A_{4}^{LC}(1,4,3,2)& 3 & 0 & 0 & 0 & 0 & 3
%     \end{array}~~~\nonumber\eea
%
%We see that coefficient of each basis matches up.
Then let us
consider non-planar basis, which come from only  the left hand side,
and we have
\bea \begin{array}{c|c|c|c|c|c|c}
   & A_4^{NP}(1,2;3,4) & A_4^{NP}(1,2;4,3) & A_4^{NP}(1,3;2,4) & A_4^{NP}(1,3;4,2) & A_4^{NP}(1,4;2,3)& A_4^{NP}(1,4;3,2) \\\hline
  A_{2,2}(1,2;3,4)& 4 & 4 & -2 & -2 & -2 & -2 \\\hline
  A_{2,2}(1,3;2,4)& -2 & -2 & 4 & 4 & -2 & -2 \\\hline
  A_{2,2}(1,4;2,3)& -2 & -2 & -2 & -2 & 4 & 4
     \end{array}~~~.\nonumber\eea
This gives a zero result, thus verifing (\ref{appendix-test-2}).

%%%%%%%%%%%%%%%%%%%%
\section{The proof of identity
$A_{1,1,2}(1;2;3,4)=A_{2,1,1}(1,2;3;4)$} %%%%%%%%%%%%%%%%

To demonstrate the use of triple cut method, we give the proof of
$A_{1,1,2}(1;2;3,4)=A_{2,1,1}(1,2;3;4)$. This identity does not
directly come from $U(1)$-decoupling equations, but is obtained from
solving these $U(1)$-decoupling equations.

Before comparing two sides under various cuts, we need to obtain
contributions of a given cut. The contributions in general are  given
by $A_L( \ell_1, \a(1), \ell_2,\b, \ell_3,\gamma)$ plus permutations
$P(\ell_1,\ell_2,\ell_3)$, where $\a(1)$ means that leg $1$
belongs to set $\a$. This is equal to fix $\ell_1$ at the first position,
but fix leg $1$ at the set $\a,\b,\gamma$ plus 
$\ell_2\leftrightarrow \ell_3$. Using this convention, we write down
contributions for the cut $s_{12}$.

For the trace structure $\mbox{Tr}(1,2) \mbox{Tr}(3) \mbox{Tr}(4)$ we
have following terms
\bea A_1 & = & [A_L(\ell_1,1,2,\ell_2,\ell_3)+\{1\leftrightarrow
2\}]\times [A_R(-\ell_1,3,-\ell_3,4, -\ell_2)+
A_R(-\ell_1,4,-\ell_3,3,-\ell_2)]~,~~~\nn
A_2 & = & [A_L(\ell_1,\ell_2,1,2,\ell_3)+\{1\leftrightarrow
2\}]\times [A_R(-\ell_1,3,-\ell_3, -\ell_2,4)+
A_R(-\ell_1,4,-\ell_3,-\ell_2,3)]~,~~~\nn
A_3 & = & [A_L(\ell_1,\ell_2,\ell_3,1,2)+\{1\leftrightarrow
2\}]\times [A_R(-\ell_1,-\ell_3, 3,-\ell_2,4)+
A_R(-\ell_1,-\ell_3,4,-\ell_2,3)]~,~~~\nn
A_4 & = & A_1(\{ \ell_2\leftrightarrow \ell_3\})~,~~~A_5  =  A_2(\{
\ell_2\leftrightarrow \ell_3\})~,~~~A_6  =  A_3(\{
\ell_2\leftrightarrow \ell_3\})~,~~~\label{A4-12-3-4}\eea
while for the trace structure $\mbox{Tr}(1)\mbox{Tr}(2)
\mbox{Tr}(3,4)$ we have following terms
\bea B_1 & = &
[A_L(\ell_1,1,\ell_2,2,\ell_3)+A_L(\ell_1,2,\ell_2,1,\ell_3)]\times[
A_R(-\ell_1,3,4,-\ell_3,-\ell_2)+\{3\leftrightarrow 4\}]~,~~~\nn
B_2 & = &
[A_L(\ell_1,1,\ell_2,\ell_3,2)+A_L(\ell_1,2,\ell_2,\ell_3,1)]\times[
A_R(-\ell_1,-\ell_3,3,4,-\ell_2)+\{3\leftrightarrow 4\}]~,~~~\nn
B_3 & = &
[A_L(\ell_1,\ell_2,1,\ell_3,2)+A_L(\ell_1,\ell_2,2,\ell_3,1)]\times[
A_R(-\ell_1,-\ell_3,-\ell_2,3,4)+\{3\leftrightarrow 4\}]~,~~~\nn
B_4 & = & B_1(\{ \ell_2\leftrightarrow \ell_3\})~,~~~B_5  =  B_2(\{
\ell_2\leftrightarrow \ell_3\})~,~~~B_6  =  B_3(\{
\ell_2\leftrightarrow \ell_3\})~.~~~\label{A4-1-2-34}\eea
To show the equality, we rewrite
\bea & & -2[A_L(\ell_1,1,2,\ell_2,\ell_3)+A_L(\ell_1,2,1,\ell_2,
\ell_3)] = A_L(\ell_1,1,\ell_2,2,\ell_3)+A_L(\ell_1, 1, \ell_2,
\ell_3, 2)\nn
& & + A_L(\ell_1,2,\ell_2,1,\ell_3)+
A_L(\ell_1,2,\ell_2,\ell_3,1)~,~~~\label{Exp-1} \eea
so the ordering with $1,2$ nearby in (\ref{A4-12-3-4}) is
transferred to the ordering with $1,2$ not nearby as given in
(\ref{A4-1-2-34}). Similarly using
\bea & &
-2[A_R(-\ell_1,3,4,-\ell_2,-\ell_3)+A_R(-\ell_1,4,3,-\ell_2,
-\ell_3)] = A_R(-\ell_1,3,-\ell_2,4,-\ell_3)+A_R(-\ell_1, 3,
-\ell_2, -\ell_3, 4)\nn
& & + A_R(-\ell_1,4,-\ell_2,3,-\ell_3)+
A_R(-\ell_1,4,-\ell_2,-\ell_3,3)~,~~~\label{Exp-2} \eea
the form in (\ref{A4-1-2-34}) will become the form  in
(\ref{A4-12-3-4}). Then we just need to put   (\ref{Exp-1})
back to (\ref{A4-12-3-4}) and (\ref{Exp-2}) back to (\ref{A4-1-2-34}),
and compare the terms in $A_i$ and $B_i$.  For example, the term with ordering
$(\ell_1,1,\ell_2,2,\ell_3)$ coming from $A_1$ and $A_2$ will be 
multiplied with
\bea {-1\over 2}[
A_R(-\ell_1,3,-\ell_3,4,-\ell_2)+A_R(-\ell_1,3,-\ell_3,
-\ell_2,4)+\{3\leftrightarrow 4\}]~,~~~\label{order-1} \eea
while the term with ordering  $(\ell_1,1,\ell_2,2,\ell_3)$ only coming from
$B_1$ will be 
 multiplied with $$A_R(-\ell_1,3,4,-\ell_3,-\ell_2) + A_R(-\ell_1,4,3,-\ell_3,-\ell_2)~,$$ which is
nothing but (\ref{order-1})  according to (\ref{Exp-2}). Other terms can
easily be checked  using the same procedure.

Cut $s_{13}$ and $s_{14}$ can be checked in the similar way, thus we 
prove the identity by unitarity cut method.

%%%%%%%%%%%%%%%%%%%%%%%%%%%%%%%%%%%%%%%%%%%%%%%%%%%%

%%%%%%%%%%%%%%%%%

\begin{thebibliography}{999}
%%%%%%%%%%%%%%%%%%%%

%%%%%%%%%%%%%%%%%%%%%%%%%%KK relation%%%%%%%%%%%%

%%%%%%%%%%%%%%%%%%%%%%%%%%%%
%Unitarity cut


%\cite{Landau:1959fi}
\bibitem{Landau:1959fi}
  L.~D.~Landau,
  %``On analytic properties of vertex parts in quantum field theory,''
  Nucl.\ Phys.\  {\bf 13}, 181 (1959);
  %%CITATION = NUPHA,13,181;%%
%\cite{Mandelstam:1958xc}
%\bibitem{Mandelstam:1958xc}

  S.~Mandelstam,
  % ``Determination of the pion - nucleon scattering amplitude from dispersion
  %relations and unitarity. General theory,''
  Phys.\ Rev.\  {\bf 112}, 1344 (1958);
  %%CITATION = PHRVA,112,1344;%%
%\cite{Mandelstam:1959bc}
%\bibitem{Mandelstam:1959bc}

  S.~Mandelstam,
  %``Analytic properties of transition amplitudes in perturbation theory,''
  Phys.\ Rev.\  {\bf 115}, 1741 (1959);
  %%CITATION = PHRVA,115,1741;%%
%\cite{Cutkosky:1960sp}
%\bibitem{Cutkosky:1960sp}

  R.~E.~Cutkosky,
  %``Singularities and discontinuities of Feynman amplitudes,''
  J.\ Math.\ Phys.\  {\bf 1}, 429 (1960).
  %%CITATION = JMAPA,1,429;%%

%\cite{Bern:1994cg}
\bibitem{Bern:1994cg}
  Z.~Bern, L.~J.~Dixon, D.~C.~Dunbar and D.~A.~Kosower,
  %``Fusing gauge theory tree amplitudes into loop amplitudes,''
  Nucl.\ Phys.\  B {\bf 435}, 59 (1995)
  [arXiv:hep-ph/9409265].
  %%CITATION = NUPHA,B435,59;%%

%\cite{Bern:1994zx}
\bibitem{Bern:1994zx}
  Z.~Bern, L.~J.~Dixon, D.~C.~Dunbar and D.~A.~Kosower,
  %``One-Loop n-Point Gauge Theory Amplitudes, Unitarity and Collinear Limits,''
  Nucl.\ Phys.\  B {\bf 425}, 217 (1994)
  [arXiv:hep-ph/9403226].
  %%CITATION = NUPHA,B425,217;%%


%\cite{Britto:2004nc}
\bibitem{Britto:2004nc}
  R.~Britto, F.~Cachazo and B.~Feng,
  %``Generalized unitarity and one-loop amplitudes in N = 4  super-Yang-Mills,''
  Nucl.\ Phys.\  B {\bf 725}, 275 (2005)
  [arXiv:hep-th/0412103].
  %%CITATION = NUPHA,B725,275;%%


%\cite{Britto:2005ha}
\bibitem{Britto:2005ha}
  R.~Britto, E.~Buchbinder, F.~Cachazo and B.~Feng,
  %``One-loop amplitudes of gluons in SQCD,''
  Phys.\ Rev.\  D {\bf 72}, 065012 (2005)
  [arXiv:hep-ph/0503132].
  %%CITATION = PHRVA,D72,065012;%%

%\cite{Anastasiou:2006jv}
\bibitem{Anastasiou:2006jv}
  C.~Anastasiou, R.~Britto, B.~Feng, Z.~Kunszt and P.~Mastrolia,
  %``D-dimensional unitarity cut method,''
  Phys.\ Lett.\  B {\bf 645}, 213 (2007)
  [arXiv:hep-ph/0609191].
  %%CITATION = PHLTA,B645,213;%%
%\cite{Anastasiou:2006gt}
%\bibitem{Anastasiou:2006gt}
  C.~Anastasiou, R.~Britto, B.~Feng, Z.~Kunszt and P.~Mastrolia,
  %``Unitarity cuts and Reduction to master integrals in d dimensions for
  %one-loop amplitudes,''
  JHEP {\bf 0703}, 111 (2007)
  [arXiv:hep-ph/0612277].
  %%CITATION = JHEPA,0703,111;%%


%%%%%%%%%%%%%%%%%%%%%%%
%CSW rules

%\cite{Cachazo:2004kj}
\bibitem{Cachazo:2004kj}
  F.~Cachazo, P.~Svrcek and E.~Witten,
  %``MHV vertices and tree amplitudes in gauge theory,''
  JHEP {\bf 0409}, 006 (2004)
  [arXiv:hep-th/0403047].
  %%CITATION = JHEPA,0409,006;%%

%\cite{Witten:2003nn}
\bibitem{Witten:2003nn}
  E.~Witten,
  %``Perturbative gauge theory as a string theory in twistor space,''
  Commun.\ Math.\ Phys.\  {\bf 252}, 189 (2004)
  [arXiv:hep-th/0312171].
  %%CITATION = CMPHA,252,189;%%




%%%%%%%%%%%%%%%%%%%%%%%%
%BCFW recursion relation

%\cite{Britto:2004ap}
\bibitem{Britto:2004ap}
  R.~Britto, F.~Cachazo and B.~Feng,
  %``New recursion relations for tree amplitudes of gluons,''
  Nucl.\ Phys.\  B {\bf 715}, 499 (2005)
  [arXiv:hep-th/0412308];
  %%CITATION = NUPHA,B715,499;%%

%\cite{Britto:2005fq}
\bibitem{Britto:2005fq}
R.~Britto, F.~Cachazo, B.~Feng and E.~Witten,
  %``Direct proof of tree-level recursion relation in Yang-Mills theory,''
  Phys.\ Rev.\ Lett.\  {\bf 94}, 181602 (2005)
  [arXiv:hep-th/0501052].
  %%CITATION = PRLTA,94,181602;%%



%\cite{Bianchi:2008pu}
%%\bibitem{Bianchi:2008pu}
%%  M.~Bianchi, H.~Elvang and D.~Z.~Freedman,
  %``Generating Tree Amplitudes in N=4 SYM and N = 8 SG,''
%%  JHEP {\bf 0809}, 063 (2008)
%%  [arXiv:0805.0757 [hep-th]].
  %%CITATION = JHEPA,0809,063;%%


%\cite{Brandhuber:2008pf}
%%\bibitem{Brandhuber:2008pf}
%%  A.~Brandhuber, P.~Heslop and G.~Travaglini,
  %``A note on dual superconformal symmetry of the N=4 super Yang-Mills
  %S-matrix,''
%%Phys.\ Rev.\  D {\bf 78}, 125005 (2008)
 %% [arXiv:0807.4097 [hep-th]].
  %%CITATION = PHRVA,D78,125005;%%

%\cite{ArkaniHamed:2008gz}
%%\bibitem{ArkaniHamed:2008gz}
%%  N.~Arkani-Hamed, F.~Cachazo and J.~Kaplan,
  %``What is the Simplest Quantum Field Theory?,''
%%  arXiv:0808.1446 [hep-th].
  %%CITATION = ARXIV:0808.1446;%%

%\cite{Elvang:2008na}
%%\bibitem{Elvang:2008na}
 %% H.~Elvang, D.~Z.~Freedman and M.~Kiermaier,
  %``Recursion Relations, Generating Functions, and Unitarity Sums in N=4 SYM
  %Theory,''
 %% JHEP {\bf 0904}, 009 (2009)
 %% [arXiv:0808.1720 [hep-th]].
  %%CITATION = JHEPA,0904,009;%%



%\cite{Drummond:2008cr}
%%\bibitem{Drummond:2008cr}
 %% J.~M.~Drummond and J.~M.~Henn,
  %``All tree-level amplitudes in N=4 SYM,''
 %% JHEP {\bf 0904}, 018 (2009)
  %%[arXiv:0808.2475 [hep-th]].
  %%CITATION = JHEPA,0904,018;%%


%\cite{Mangano:1987xk}
\bibitem{Mangano:1987xk}
  M.~L.~Mangano, S.~J.~Parke and Z.~Xu,
  %``Duality and Multi - Gluon Scattering,''
  Nucl.\ Phys.\  B {\bf 298}, 653 (1988).
  %%CITATION = NUPHA,B298,653;%%

%\cite{Bern:1990ux}
\bibitem{Bern:1990ux}
  Z.~Bern and D.~A.~Kosower,
  %``Color decomposition of one loop amplitudes in gauge theories,''
  Nucl.\ Phys.\  B {\bf 362}, 389 (1991).
  %%CITATION = NUPHA,B362,389;%%


%\cite{Mangano:1990by}
\bibitem{Mangano:1990by}
  M.~L.~Mangano and S.~J.~Parke,
  %``Multiparton amplitudes in gauge theories,''
  Phys.\ Rept.\  {\bf 200}, 301 (1991)
  [arXiv:hep-th/0509223].
  %%CITATION = PRPLC,200,301;%%







%\cite{Kleiss:1988ne}
\bibitem{Kleiss:1988ne}
  R.~Kleiss and H.~Kuijf,
  %``MULTI - GLUON CROSS-SECTIONS AND FIVE JET PRODUCTION AT HADRON COLLIDERS,''
  Nucl.\ Phys.\  B {\bf 312}, 616 (1989).
  %%CITATION = NUPHA,B312,616;%%


%\cite{DelDuca:1999rs}
\bibitem{DelDuca:1999rs}
  V.~Del Duca, L.~J.~Dixon and F.~Maltoni,
  %``New color decompositions for gauge amplitudes at tree and loop level,''
  Nucl.\ Phys.\  B {\bf 571}, 51 (2000)
  [arXiv:hep-ph/9910563].
  %%CITATION = NUPHA,B571,51;%%

%\cite{BjerrumBohr:2009rd}
\bibitem{BjerrumBohr:2009rd}
  N.~E.~J.~Bjerrum-Bohr, P.~H.~Damgaard and P.~Vanhove,
  %``Minimal Basis for Gauge Theory Amplitudes,''
  Phys.\ Rev.\ Lett.\  {\bf 103}, 161602 (2009)
  [arXiv:0907.1425 [hep-th]].
  %%CITATION = PRLTA,103,161602;%%

%\cite{Stieberger:2009hq}
\bibitem{Stieberger:2009hq}
  S.~Stieberger,
  %``Open & Closed vs. Pure Open String Disk Amplitudes,''
  arXiv:0907.2211 [hep-th].
  %%CITATION = ARXIV:0907.2211;%%




%\cite{Bern:2008qj}
\bibitem{Bern:2008qj}
  Z.~Bern, J.~J.~M.~Carrasco and H.~Johansson,
  %``New Relations for Gauge-Theory Amplitudes,''
  Phys.\ Rev.\  D {\bf 78}, 085011 (2008)
  [arXiv:0805.3993 [hep-ph]].
  %%CITATION = PHRVA,D78,085011;%%

%\cite{Feng:2010my}
\bibitem{Feng:2010my}
  B.~Feng, R.~Huang and Y.~Jia,
  %``Gauge Amplitude Identities by On-shell Recursion Relation in S-matrix
  %Program,''
  arXiv:1004.3417 [hep-th].
  %%CITATION = ARXIV:1004.3417;%%

%\cite{Chen:2011jx}
\bibitem{Chen:2011jx}
  Y.~X.~Chen, Y.~J.~Du and B.~Feng,
  %``A Proof of the Explicit Minimal-basis Expansion of Tree Amplitudes in Gauge
  %Field Theory,''
  JHEP {\bf 1102}, 112 (2011)
  [arXiv:1101.0009 [hep-th]].
  %%CITATION = JHEPA,1102,112;%%


%\cite{Jia:2010nz}
\bibitem{Jia:2010nz}
  Y.~Jia, R.~Huang and C.~Y.~Liu,
  %``U(1)-decoupling, KK and BCJ relations in $\mathcal{N}=4$ SYM,''
  Phys.\ Rev.\  D {\bf 82}, 065001 (2010)
  [arXiv:1005.1821 [hep-th]].
  %%CITATION = PHRVA,D82,065001;%%


%\cite{Huang:2010fc}
\bibitem{Huang:2010fc}
  J.~H.~Huang, R.~Huang and Y.~Jia,
  %``Tree amplitudes of noncommutative U(N) Yang-Mills Theory,''
  arXiv:1009.5073 [hep-th].
  %%CITATION = ARXIV:1009.5073;%%

%\cite{Zhang:2010ve}
\bibitem{Zhang:2010ve}
  H.~Zhang,
  %``Note on the non-adjacent BCFW deformations,''
  arXiv:1005.4462 [hep-th].
  %%CITATION = ARXIV:1005.4462;%%



%\cite{Berends:1987cv}
\bibitem{Berends:1987cv}
  F.~A.~Berends and W.~Giele,
  %``The Six Gluon Process As An Example Of Weyl-Van Der Waerden Spinor
  %Calculus,''
  Nucl.\ Phys.\  B {\bf 294}, 700 (1987).
  %%CITATION = NUPHA,B294,700;%%



%\cite{Mangano:1988kk}
\bibitem{Mangano:1988kk}
  M.~L.~Mangano,
  %``The Color Structure Of Gluon Emission,''
  Nucl.\ Phys.\  B {\bf 309}, 461 (1988).
  %%CITATION = NUPHA,B309,461;%%

%\cite{Dixon:1996wi}
\bibitem{Dixon:1996wi}
  L.~J.~Dixon,
  %``Calculating scattering amplitudes efficiently,''
  arXiv:hep-ph/9601359.
  %%CITATION = HEP-PH/9601359;%%



%\cite{Sondergaard:2009za}
\bibitem{Sondergaard:2009za}
  T.~Sondergaard,
  %``New Relations for Gauge-Theory Amplitudes with Matter,''
  Nucl.\ Phys.\  B {\bf 821}, 417 (2009)
  [arXiv:0903.5453 [hep-th]].
  %%CITATION = NUPHA,B821,417;%%









%\cite{Tye:2010dd}
\bibitem{Tye:2010dd}
  S.~H.~Henry Tye and Y.~Zhang,
  %``Dual Identities inside the Gluon and the Graviton Scattering Amplitudes,''
  JHEP {\bf 1006}, 071 (2010)
  [arXiv:1003.1732 [hep-th]].
  %%CITATION = JHEPA,1006,071;%%

%\cite{BjerrumBohr:2010zs}
\bibitem{BjerrumBohr:2010zs}
  N.~E.~J.~Bjerrum-Bohr, P.~H.~Damgaard, T.~Sondergaard and P.~Vanhove,
  %``Monodromy and Jacobi-like Relations for Color-Ordered Amplitudes,''
  JHEP {\bf 1006}, 003 (2010)
  [arXiv:1003.2403 [hep-th]].
  %%CITATION = JHEPA,1006,003;%%

%\cite{Mafra:2009bz}
\bibitem{Mafra:2009bz}
  C.~R.~Mafra,
  %``Simplifying the Tree-level Superstring Massless Five-point Amplitude,''
  JHEP {\bf 1001}, 007 (2010)
  [arXiv:0909.5206 [hep-th]].
  %%CITATION = JHEPA,1001,007;%%

%\cite{Mafra:2010jq}
\bibitem{Mafra:2010jq}
  C.~R.~Mafra, O.~Schlotterer, S.~Stieberger and D.~Tsimpis,
  %``A recursive formula for N-point SYM tree amplitudes,''
  arXiv:1012.3981 [hep-th].
  %%CITATION = ARXIV:1012.3981;%%

%\cite{BjerrumBohr:2011xe}
\bibitem{BjerrumBohr:2011xe}
  N.~E.~J.~Bjerrum-Bohr, P.~H.~Damgaard, H.~Johansson and T.~Sondergaard,
  %``Monodromy--like Relations for Finite Loop Amplitudes,''
  arXiv:1103.6190 [hep-th].
  %%CITATION = ARXIV:1103.6190;%%









%\cite{Buchbinder:2005wp}
\bibitem{Buchbinder:2005wp}
  E.~I.~Buchbinder and F.~Cachazo,
  %``Two-loop amplitudes of gluons and octa-cuts in N=4 super Yang-Mills,''
  JHEP {\bf 0511}, 036 (2005)
  [arXiv:hep-th/0506126].
  %%CITATION = JHEPA,0511,036;%%

%\cite{Cachazo:2008vp}
\bibitem{Cachazo:2008vp}
  F.~Cachazo,
  %``Sharpening The Leading Singularity,''
  arXiv:0803.1988 [hep-th].
  %%CITATION = ARXIV:0803.1988;%%


%\cite{Cachazo:2008hp}
\bibitem{Cachazo:2008hp}
  F.~Cachazo, M.~Spradlin and A.~Volovich,
  %``Leading Singularities of the Two-Loop Six-Particle MHV Amplitude,''
  Phys.\ Rev.\  D {\bf 78}, 105022 (2008)
  [arXiv:0805.4832 [hep-th]].
  %%CITATION = PHRVA,D78,105022;%%




%\cite{ArkaniHamed:2009dn}
\bibitem{ArkaniHamed:2009dn}
  N.~Arkani-Hamed, F.~Cachazo, C.~Cheung and J.~Kaplan,
  %``A Duality For The S Matrix,''
  JHEP {\bf 1003}, 020 (2010)
  [arXiv:0907.5418 [hep-th]].
  %%CITATION = JHEPA,1003,020;%%


%\cite{Bern:2010ue}
\bibitem{Bern:2010ue}
  Z.~Bern, J.~J.~M.~Carrasco and H.~Johansson,
  %``Perturbative Quantum Gravity as a Double Copy of Gauge Theory,''
  Phys.\ Rev.\ Lett.\  {\bf 105}, 061602 (2010)
  [arXiv:1004.0476 [hep-th]].
  %%CITATION = PRLTA,105,061602;%%


%\cite{Bern:2010yg}
\bibitem{Bern:2010yg}
  Z.~Bern, T.~Dennen, Y.~t.~Huang and M.~Kiermaier,
  %``Gravity as the Square of Gauge Theory,''
  Phys.\ Rev.\  D {\bf 82}, 065003 (2010)
  [arXiv:1004.0693 [hep-th]].
  %%CITATION = PHRVA,D82,065003;%%



%\cite{Bern:2011ia}% %\cite{Bern:2010ue,Bern:2010yg,BjerrumBohr:2011xe,Bern:2011ia}%
\bibitem{Bern:2011ia}
  Z.~Bern and T.~Dennen,
  %``A Color Dual Form for Gauge-Theory Amplitudes,''
  arXiv:1103.0312 [hep-th].
  %%CITATION = ARXIV:1103.0312;%%



%\cite{Carrasco:2011hw}
\bibitem{Carrasco:2011hw}
  J.~J.~M.~Carrasco and H.~Johansson,
  %``Generic multiloop methods and application to N=4 super-Yang-Mills,''
  arXiv:1103.3298 [hep-th].
  %%CITATION = ARXIV:1103.3298;%%



\bibitem{PV}
 L.M.\ Brown and R.P.\ Feynman, Phys.\ Rev.\ 85:231 (1952);

 G.\ Passarino and M.\ Veltman, Nucl.\ Phys.\ {B160:151} (1979);

 G. 't Hooft and M. Veltman, Nucl.\ Phys.\ {B153:365} (1979);

 R. G. Stuart, Comp.\ Phys.\ Comm.\ 48:367 (1988);

 R. G. Stuart and A. Gongora, Comp.\ Phys.\ Comm.\ 56:337 (1990).

%\cite{Gluza:2010ws}
\bibitem{Gluza:2010ws}
  J.~Gluza, K.~Kajda and D.~A.~Kosower,
  %``Towards a Basis for Planar Two-Loop Integrals,''
  Phys.\ Rev.\  D {\bf 83}, 045012 (2011)
  [arXiv:1009.0472 [hep-th]].
  %%CITATION = PHRVA,D83,045012;%%

%\cite{Bern:1997nh}
\bibitem{Bern:1997nh}
  Z.~Bern, J.~S.~Rozowsky and B.~Yan,
  %``Two loop four gluon amplitudes in N=4 superYang-Mills,''
  Phys.\ Lett.\  B {\bf 401}, 273 (1997)
  [arXiv:hep-ph/9702424].
  %%CITATION = PHLTA,B401,273;%%

%\cite{Luo:2004ss}
\bibitem{Luo:2004ss}
  M.~x.~Luo and C.~k.~Wen,
  %``One-loop maximal helicity violating amplitudes in N=4 super Yang-Mills
  %theories,''
  JHEP {\bf 0411}, 004 (2004)
  [arXiv:hep-th/0410045].
  %%CITATION = JHEPA,0411,004;%%


%\cite{Luo:2004nw}
\bibitem{Luo:2004nw}
  M.~x.~Luo and C.~k.~Wen,
  %``Systematics of one-loop scattering amplitudes in N=4 super Yang-Mills
  %theories,''
  Phys.\ Lett.\  B {\bf 609}, 86 (2005)
  [arXiv:hep-th/0410118].
  %%CITATION = PHLTA,B609,86;%%

%\cite{Bern:2000dn}
\bibitem{Bern:2000dn}
  Z.~Bern, L.~J.~Dixon and D.~A.~Kosower,
  %``A Two loop four gluon helicity amplitude in QCD,''
  JHEP {\bf 0001}, 027 (2000)
  [arXiv:hep-ph/0001001].
  %%CITATION = JHEPA,0001,027;%%


%\cite{Bern:2008ap}
\bibitem{Bern:2008ap}
  Z.~Bern, L.~J.~Dixon, D.~A.~Kosower, R.~Roiban, M.~Spradlin, C.~Vergu and A.~Volovich,
  %``The Two-Loop Six-Gluon MHV Amplitude in Maximally Supersymmetric Yang-Mills
  %Theory,''
  Phys.\ Rev.\  D {\bf 78}, 045007 (2008)
  [arXiv:0803.1465 [hep-th]].
  %%CITATION = PHRVA,D78,045007;%%

%\cite{Drummond:2010mb}
\bibitem{Drummond:2010mb}
  J.~M.~Drummond and J.~M.~Henn,
  %``Simple loop integrals and amplitudes in N=4 SYM,''
  JHEP {\bf 1105}, 105 (2011)
  [arXiv:1008.2965 [hep-th]].
  %%CITATION = JHEPA,1105,105;%%

%\cite{Gluza:2010ws}
\bibitem{Gluza:2010ws}
  J.~Gluza, K.~Kajda and D.~A.~Kosower,
  %``Towards a Basis for Planar Two-Loop Integrals,''
  Phys.\ Rev.\  D {\bf 83}, 045012 (2011)
  [arXiv:1009.0472 [hep-th]].
  %%CITATION = PHRVA,D83,045012;%%

%\cite{Bern:2002tk}
\bibitem{Bern:2002tk}
  Z.~Bern, A.~De Freitas and L.~J.~Dixon,
  %``Two loop helicity amplitudes for gluon-gluon scattering in QCD and
  %supersymmetric Yang-Mills theory,''
  JHEP {\bf 0203}, 018 (2002)
  [arXiv:hep-ph/0201161].
  %%CITATION = JHEPA,0203,018;%%

%\cite{Bern:2010tq}
\bibitem{Bern:2010tq}
  Z.~Bern, J.~J.~M.~Carrasco, L.~J.~Dixon, H.~Johansson and R.~Roiban,
  %``The Complete Four-Loop Four-Point Amplitude in N=4 Super-Yang-Mills
  %Theory,''
  Phys.\ Rev.\  D {\bf 82}, 125040 (2010)
  [arXiv:1008.3327 [hep-th]].
  %%CITATION = PHRVA,D82,125040;%%


\end{thebibliography}
\end{document}